\documentclass[aps,prb,reprint,groupedaddress, footinbib]{revtex4-1}
\usepackage{amsmath}
\usepackage{graphicx}
\usepackage{hyperref}
\usepackage{bbold}
\usepackage{color}
\begin{document}
	\title{Weak localization in transition metal dichalcogenide monolayers and their heterostructures with graphene}
	\author{Stefan Ili\'{c}, Julia S. Meyer, and Manuel Houzet}
	\affiliation{Univ.~Grenoble Alpes, CEA, INAC-Pheliqs, F-38000 Grenoble, France}
	\date{\today}
	\begin{abstract}
		We calculate the interference correction to the conductivity of doped transition metal dichalcogenide \textcolor{black}{(TMDC)} monolayers. Because of the interplay between valley structure and intrinsic spin-orbit coupling (SOC), these materials exhibit a rich weak localization (WL) behavior that is qualitatively different from conventional metals or similar two-dimensional materials such as graphene. Our results can also be used to describe graphene/TMDC heterostructures, where the SOC is induced in the graphene sheet. We discuss new parameter regimes that go beyond existing theories, and can be used to interpret recent experiments in order to assess the strength of SOC and disorder. Furthermore, we show that an in-plane Zeeman field can be used to distinguish the contributions  of different kinds of SOC to the WL magneto\textcolor{black}{conductance}.

	\end{abstract}
	\maketitle
	\section{Introduction}
Transition metal dichalcogenide \textcolor{black}{(TMDC)} monolayers are a class of two-dimensional semiconductors of the form MX$_2$, where M is a transition metal and X is a chalcogen. Similarly to graphene, TMDCs have a hexagonal lattice structure, and a number of them (M=Mo, W; X=S, Se, Te) have minima/maxima of the conduction/valence band at the two corners (valleys)  $\pm \mathbf{K}$ of the Brillouin zone. Unlike graphene, however, TMDCs have two inequivalent lattice sites and no inversion symmetry, which allows for a large band gap in their spectrum \cite{mak,wang}.

 Because of the heavy constituent atoms, these materials also host strong intrinsic spin-orbit coupling (SOC), which acts as an effective out-of-plane Zeeman field with opposite orientation in the two valleys \cite{zhu2011giant, kormanyos2015k, xiao}. This valley-dependent SOC enables a variety of applications of TMDCs in optoelectronics and so-called valleytronics, as electrons from different valleys can be excited selectively with circularly polarized light \cite{mak2012control, zeng2012valley}. When sufficiently doped, several TMDCs become superconducting \cite{saito, lu, xi2015ising,de2017tuning}, where intrinsic SOC plays an important role, as it causes unconventional \lq\lq Ising pairing\rq\rq\, of the Cooper pairs and a great enhancement of the in-plane upper critical field \cite{saito,ilic2017enhancement}.

The possibility of inducing SOC in a graphene sheet by coupling it to \textcolor{black}{gapped} TMDCs in heterostructures has recently sparked scientific interest, as it can lead to phenomena such as edge states \cite{gmitra2016trivial, yang2016tunable} and  the spin Hall effect \textcolor{black}{\cite{avsar2014spin,garcia2017spin,milletari2017covariant}}. The induced SOC originates from hybridization of the transition metal and carbon orbitals \cite{yang2016tunable}. It has two contributions: Kane-Mele SOC \cite{kane2005quantum}, which can open a topological gap at the Dirac points $\pm\mathbf{K}$, and so-called valley-Zeeman SOC, which breaks the inversion symmetry of graphene and causes \textcolor{black}{spin splitting} in the band structure.

Transport measurements in \textcolor{black}{doped} TMDCs \cite{schmidt2016quantum, zhang2017robustly,costanzo2016gate} and graphene/TMDC  heterostructures \cite{wang2016origin, yang2017strong,volkl2017magnetotransport, wakamura2017strong, zihlmann2017large, yang2016tunable} can give information about the amplitude and mechanism of SOC by studying the quantum correction to the conductance, due to weak localization (WL) and/or antilocalization (WAL) of electrons. W(A)L can be probed by applying a perpendicular magnetic field $B_\bot$, which suppresses the quantum correction by breaking time\textcolor{black}{-}reversal symmetry. By measuring the resulting magnetoconductance as a function of $B_\bot$ and fitting it to theoretical models, one can extract parameters such as scattering and spin relaxation rates.

 So far, the experiments have been interpreted using the so-called Hikami-Larkin-Nagaoka (HLN) \cite{hikami1980spin} formula (for TMDC experiments \textcolor{black}{\cite{schmidt2016quantum, zhang2017robustly,costanzo2016gate}})  or a similar formula provided by the McCann-Fal'ko (MF)\cite{mccann2012z} theory in the regime of strong intervalley scattering (for graphene/TMDC experiments \textcolor{black}{\cite{wang2016origin, yang2017strong,volkl2017magnetotransport, wakamura2017strong, zihlmann2017large, yang2016tunable}}).
HLN theory holds for two-dimensional single-band systems in the presence of SOC. If SOC is weak, constructive electron interference along time-reversed trajectories gives rise to a decrease in conductance (WL). \textcolor{black}{By contrast,} strong SOC leads to a phase shift due to the spin precession,  which results in destructive interference and an increase in conductance (WAL). In Dirac materials, such as TMDCs and graphene, the physical picture becomes more complex. Here, the quantum correction is sensitive to the sublattice degree of freedom, or so-called lattice isospin. Due to the associated Berry phase, it can introduce phase shifts similarly to the spin physics. Furthermore, the multivalley nature of these materials and intervalley scattering also influence the quantum correction. MF theory takes these effects into account for the case of graphene, and gives a full description of WL and WAL with any disorder that satisfies time-reversal symmetry. In the presence of spin-orbit impurities and in the regime of strong  intervalley scattering, such that the valley physics is suppressed, it reduces to the HLN formula. 

However, the applicability of MF and HLN theories to TMDC\textcolor{black}{s} and graphene/TMDC is limited, since they were both developed to describe spin-degenerate systems and do not capture \textcolor{black}{the} spin splitting caused by the presence of valley-dependent SOC. A theory for TMDCs that takes it into account was given by Ochoa \emph{et al.}~\cite{ochoa2014spin} in the regime close to the bottom/top of the conduction/valence band, $|\mu|\approx E_g$, where $\mu$ is the chemical potential and $2 E_g$ is the band-gap. This parameter regime, however, does not fully describe graphene/TMDC heterostructures and highly doped TMDCs, where $|\mu|\gg E_g$.

In this work, we  present a general theory of the interference correction for a massive Dirac material with valley-Zeeman SOC. Furthermore, we account for the effect of an in-plane Zeeman field. Our formula can be  applied to TMDC\textcolor{black}{s} and graphene/TMDC heterostructures. Namely, we generalize Ref.~\onlinecite{ochoa2014spin} to any chemical potential $\mu$, and we show that several contributions to the interference-induced magneto\textcolor{black}{conductance} are sensitive to the magnitude of doping, and are modified or suppressed as the doping increases. We discuss in detail the regime where intervalley scattering dominates over any spin-dependent scattering, which is the most commonly invoked regime when interpreting the experimental data.  We find that the interplay between valley-dependent SOC\textcolor{black}{,} $\Delta_{so}$\textcolor{black}{,} and intervalley scattering, parametrized by the scattering time $\tau_{iv}$, leads to new regimes of WL and WAL. In the limit $\tau_{iv}^{-1}\gg \Delta_{so}$, MF still holds and HLN is valid if $\tau_{iv}^{-1}\gg \tau_\phi^{-1}$, where $\tau_{\phi}$ accounts for inelastic dephasing of electrons. However, we find new behavior not captured by these formulas if $\Delta_{so} \gtrsim \tau_{iv}^{-1}$. Since both TMDCs and graphene/TMDC are expected to have substantial valley-dependent SOC\cite{kormanyos2015k,gmitra2016trivial,yang2016tunable}, our newfound regimes are experimentally relevant and can be used to extract parameters from the interference-induced magneto\textcolor{black}{conductance} in both systems.

This article is organized in the following way: In Sec.~\ref{sec2}, we introduce the model Hamiltonian for disordered TMDC\textcolor{black}{s} and graphene/TMDC heterostructures. In Sec.~\ref{sec3}, we calculate the interference correction for these materials using the standard diagrammatic technique for disordered systems. We discuss our results in Sec.~\ref{sec4}. 

\section{The model}\label{sec2}

The low-energy Hamiltonian describing TMDC monolayers in the vicinity of the $\pm \mathbf{K}$ points, and in the presence of a parallel magnetic field is given by \cite{kormanyos2015k} $H_\mathbf{q}=H_0+H_{SOC}+H_W+H_{||}$, where
\begin{align}
H_0= & \, v(q_x\sigma_x\eta_z+q_y\sigma_y)+E_g\sigma_z, \nonumber \\
H_{SOC}= & \, \Delta_{KM} \sigma_zs_z\eta_z+ \Delta_{VZ} s_z\eta_z+
\lambda (\sigma_xs_y\eta_z-\sigma_y s_x) \nonumber \\ 
&+\zeta (q_x\sigma_xs_z+q_y\sigma_y  s_z\eta_z), \nonumber  \\
H_W= & \,\kappa (q_x^2-q_y^2)\sigma_x-2\kappa q_x q_y \sigma_y\eta_z,  \nonumber \\
H_{||}= & \, h s_x.
\label{eqn1}
\end{align}
Here, we use units where $\hbar=k_B=1$. The two Dirac cones are described by $H_0$, where $\mathbf{q}=(q_x,q_y)=q(\cos \theta, \sin \theta)$ is a small momentum measured from $\pm \mathbf{K}$, $v$ is the velocity associated with the linearized kinetic dispersion, and $E_g$ is the difference in on-site energy responsible for opening the band gap. Spin-orbit coupling is described by $H_{SOC}$, where $\Delta_{KM}$ and $\Delta_{VZ}$ characterize Kane-Mele and valley-Zeeman SOC, respectively. Rashba SOC, which is related to a mirror symmetry breaking due to the substrate or external fields, is described by $\lambda$. The spin-dependence of the velocity is accounted for by $\zeta$. $H_W$ describes the so-called trigonal warping. Finally, $H_{||}$ is the in-plane Zeeman field, where the Zeeman energy $h=\frac{1}{2}g\mu_B B_{||}$ is determined by the amplitude of the in-plane magnetic field and the $g$-factor, which is expected to take the value  $g\approx 2$ in these materials. We introduce Pauli matrices $\sigma_{x,y,z}$, $s_{x,y,z}$ and $\eta_{x,y,z}$  acting in sublattice, spin, and valley space, respectively. The Hamiltonian \eqref{eqn1} contains all terms up to the first order in $\mathbf{q}$ allowed by the symmetries of the system, as well as $H_W$ and $H_{||}$, which break rotational and time-reversal symmetry, respectively.

Furthermore, the low-energy sector of graphene/TMDC heterostructures is also well described by the Hamiltonian \eqref{eqn1}. First-principle calculations \cite{gmitra2016trivial,yang2016tunable} show that the Dirac cones of graphene in these heterostructures are preserved and are within the TMDC band gap. The coupling to the TMDC modifies the graphene spectrum by introducing the staggered sublattice potential, $E_g\sigma_z$, and SOC, $H_{SOC}$. 

\textcolor{black}{To proceed,} we assume that the Dirac Hamiltonian $H_0$ gives the dominant contribution to the energy of the system. $H_0$ is diagonalized by a unitary transformation
$U_\mathbf{q}=e^{-i\eta_z\alpha_\mathbf{q}}\,e^{i\beta_\mathbf{q}\sigma_y\eta_z}\,e^{i\alpha_\mathbf{q}\sigma_z\eta_z}$, where $\tan(2 \alpha_\mathbf{q})=q_y/q_x$ and $\tan(2 \beta_\mathbf{q})=vq/E_g$. It has a simple spectrum, $\pm E_{\mathbf{q}}=\pm\sqrt{q^2v^2+E_g^2}$. After projecting $U_\mathbf{q} H_{\mathbf{q}}U_{\mathbf{q}}^{\dagger}$ onto the conduction band, we obtain the effective Hamiltonian
\begin{align}
\mathcal{H}_{\mathbf{q}}=&\,\xi_{\mathbf{q}}+\Delta_{so}s_z\eta_z+\lambda\frac{vq_F}{\mu}(s_y\cos\theta-s_x\sin\theta) \nonumber  \\
&+\kappa \frac{v q_F^3}{\mu}\cos 3\theta \,\eta_z+h s_x.
\label{eqn2}
\end{align}
Here, the energy is measured from the chemical potential, $\xi_{\mathbf{q}}=E_{\mathbf{q}}-\mu$. Furthermore, we have introduced the Fermi momentum $q_F=\sqrt{\mu^2-E_g^2}/v$ and \textcolor{black}{spin-orbit splitting} $\Delta_{so}=\Delta_{KM} E_g/\mu+\Delta_{VZ}+\zeta vq_F^2/\mu$. \textcolor{black}{Note that at $\mu\gg E_g$ (as in the case of graphene, e.g.), Kane-Mele SOC does not contribute to the spin-orbit splitting.} The chemical potential $\mu$ is assumed to be sufficiently above the band gap $E_g$, so that it is the dominant energy scale, $\textcolor{black}{|\mu|}-E_g\gg \Delta_{so},\lambda, h, \kappa q_F^2$. A Hamiltonian of a similar form can be found in the valence band after the substitution $\xi_\mathbf{q}\rightarrow -\xi_{\mathbf{q}}$, $\mu\rightarrow -\mu$. Although, in the remainder of the text, we will focus only on the conduction band for simplicity, our results also hold in the valence band as long as both spin-split band\textcolor{black}{s} are occupied. This is readily  achieved in graphene/TMDC heterostructures, while a very high doping is required in TMDCs, due to the large spin-splitting caused by the intrinsic SOC in the valence band \cite{kormanyos2015k}.

 The effect of \textcolor{black}{potential} impurities can be modeled by introducing a random disorder, $H^{D0}_{\mathbf{qq'}}=U^0_{\mathbf{q-q'}}+\textcolor{black}{\sum_{i=\pm,x}\sum_{j=x,y}V^{ij}_{\mathbf{q-q'}}\sigma_i\eta_j}$,\textcolor{black}{where $\sigma_{\pm}=1\pm \sigma_z$. The first term is the intravalley contribution, which is diagonal in spin and sublattice space. The second term represents all spin-independent intervalley contributions allowed by time-reversal  \footnote{ Note that the time-reversal operator in this basis is $\mathcal{T}=i s_y\eta_x \mathcal{K}$, where $\mathcal{K}$ is complex conjugation.}  and $x-y$ symmetry. Intervalley disorder requires large momentum transfer, and is caused by short-range impurities, such as atomic defects}. Upon rotating  $U_\mathbf{q}H_{\mathbf{qq'}}^{D0}U^\dagger_{\mathbf{q'}}$ and projecting to the conduction band, a variety of other scattering processes will be generated as combinations of the band structure and potential scattering parameters.

 For simplicity, we will account for these processes, \textcolor{black}{as well as all other possible scattering processes}, phenomenologically, by independent scattering potentials. To do so, we supplement $H^{D0}_{\mathbf{qq'}}$ with all the other disorder terms allowed by the time-reversal symmetry, as was done previously in similar studies of weak localization \cite{mccann2012z, ochoa2014spin}. The disorder Hamiltonian is then given as $H^{D}_{\mathbf{qq'}}=H^{D0}_{\mathbf{qq'}}+\delta H^D_{\mathbf{qq'}}$, where
 \begin{align}
 \delta H^D_{\mathbf{qq'}}=&\sum_{i=x,y,z} U^{i}_{\mathbf{q-q'}}\Sigma_i+\sum_{i=0,x,y,z}\sum_{j=x,y,z}A^{ij}_{\mathbf{q-q'}}\Sigma_i  s_j \eta_z\nonumber \\
 +& \textcolor{black}{\sum_{j=x,y}
\sum_{i=x,y,z}M^{ij}_{\mathbf{q-q'}} \sigma_y s_i\eta_j.}
 \label{eqn3}
 \end{align}
 Here $\Sigma_{0,z,x}=\sigma_{0,x,z}$ and $\Sigma_y=\sigma_y\eta_z$. \textcolor{black}{The first line in Eq.~\eqref{eqn3} describes intravalley disorder. Here, the first and second term account for spin-dependent and spin-independent contributions, respectively. The second line describes spin-dependent intervalley disorder.} We characterize the random disorder potentials by Gaussian correlators and assume that different kinds of disorder are uncorrelated:
 \begin{align}
 &\langle U^i_{\mathbf{q}}U^j_{\mathbf{q'}}\rangle=U_i^2\delta_{ij}\delta_{\mathbf{q\bar{q}'}}, \nonumber \\
 &\langle X^{ij}_{\mathbf{q}}X^{kl}_{\mathbf{q'}}\rangle=X_{ij}^2 \delta_{ik}\delta_{jl}\delta_{\mathbf{q\bar{q}'}}.
 \label{eqn4}
 \end{align}
 Here, the brackets $\langle...\rangle$ represent disorder averaging and $X=A,V,M$. Furthermore, we use the abbreviation $\mathbf{\bar{q}}=-\mathbf{q}$.
  
We proceed by writing the rotated phenomenological disorder potential, $U_\mathbf{q}H^D_{\mathbf{qq'}}U^\dagger_\mathbf{q'}$, in the projected basis
\begin{align}
&\mathcal{H}^D_{\mathbf{qq'}}=\sum_{i=0,x,y,z}\bigg[U^i_{\mathbf{q-q'}} f^i_{\theta,\theta'}+
 \sum_{j=x,y,z}A^{ij}_{\mathbf{q-q'}}f^{i}_{\theta,\theta'} s_j \eta_z\bigg]  \nonumber \\
&+\sum_{j=x,y}\bigg[\sum_{i=\textcolor{black}{\pm,x}}V^{ij}_{\mathbf{q-q'}}g^i_{\theta,\theta'}\eta_j +
\sum_{i=x,y,z}M^{ij}_{\mathbf{q-q'}}g^y_{\theta,\theta'} s_i \eta_j\bigg],
\label{eqn5}
\end{align}
where the functions $f^i_{\theta,\theta'}$ and $g^i_{\theta\theta'}$ capture the anisotropy of the projected disorder potential, which is due to the momentum dependence of the unitary transformation $U_\mathbf{q}$. In particular, $2f^0_{\theta,\theta'}=1+e^{-i\eta_z(\theta-\theta')}+\frac{E_g}{\mu}\big(1-e^{-i\eta_z(\theta-\theta')}\big)$ and
 $2f^x_{\theta,\theta'}=\frac{vq_F}{\mu}\big(e^{-i\eta_z\theta}+e^{i\eta_z\theta'}\big)\eta_z$. Furthermore,  $f^y_{\theta,\theta'}=i f^x_{\theta,\bar{\theta}'}\eta_z$, $f^z_{\theta,\theta'}=f^0_{\bar{\theta},\theta'}$, $\textcolor{black}{g^+_{\theta,\theta'}=(1+\frac{E_g}{\mu})}$, $\textcolor{black}{g^-_{\theta,\theta'}=(\frac{E_g}{\mu}-1)e^{i\eta_z(\theta+\theta')}}$, $g^x_{\theta,\theta'}=f^0_{-\bar{\theta},\theta'}$, and $g^y_{\theta,\theta'}=i\eta_z f^x_{-\theta,\theta'}$. Here, we used the notation $\bar{\theta}=\theta+\pi$. In simple metals, anisotropic disorder usually only leads to the renormalization of the diffusion constant and the transport time. It has more profound physical consequences in our system, as it captures the sublattice \textcolor{black}{isospin} physics and the effect of the Berry curvature.

In order to describe quantum transport in our system, we will employ \textcolor{black}{the} standard diagrammatic technique for disordered systems. In particular, we introduce disorder-averaged, zero-temperature retarded $(R)$ and advanced $(A)$  Green's functions as
\begin{equation}
G^{R,A}_{\mathbf{q}\omega}=\bigg(\omega-\mathcal{H}_{\mathbf{q}}\pm \frac{i}{2\tau}\bigg)^{-1}.
\label{eqn6}
\end{equation}
Here, the self-energy $\pm i /(2\tau) $ is calculated from the self-consistent Born approximation, $\omega$ is the frequency,  and the inverse scattering time $\tau^{-1}$ is given by
\begin{equation}
\tau^{-1}=\tau_0^{-1}+\tau_{z}^{-1}+\tau_{iv}^{-1}+\sum_{i=z,zv,iv}\sum_{j={e,o}}\tau^{-1}_{i,j}.
\label{eqn7}
\end{equation}
The individual contributions to Eq.~\eqref{eqn7} are defined in the left column of Table~\ref{table1}, where we introduced the Fermi velocity\textcolor{black}{,} $v_F=v^2q_F/\mu$\textcolor{black}{,} and the density of states per valley and per spin at the Fermi level\textcolor{black}{,} $\nu=\mu/(2\pi v_F^2)$. Furthermore, we will  assume that the diagonal disorder rate $\tau_0^{-1}$ is the dominant one, i.e., $\tau^{-1}\approx \tau_0^{-1}$, and we will use the diffusive approximation $\textcolor{black}{|\mu|}-E_g\gg \tau_0^{-1}\gg \Delta_{so}, h, \lambda, \kappa q_{\textcolor{black}{F}}^2.$ 
 
 \begin{table*}[]
 	\begin{tabular}{p{4.6cm}p{4.7cm}p{2.8cm}| p{5cm} }
 		\hline \hline
 		\textcolor{black}{Intravalley scattering rates} & & & Estimates \\
 		\hline
 		$\tau_0^{-1}=\pi \nu U_0^2(1+\frac{E_g^2}{\mu^2})$ & & & $\qquad$ /\\
 		$\tau_{z1}^{-1}=\pi\nu( U_x^2+U_y^2)\frac{v^2q_F^2}{\mu^2}$  & $ \tau_{z2}^{-1}=\pi\nu U_z^2(1+\frac{E_g^2}{\mu^2})$ 
 		& $\tau_z^{-1}=\tau_{z1}^{-1}+\tau_{z2}^{-1}$  & $\textcolor{black}{\tau_{z1}^{-1},\tau_{z2}^{-1} \propto \tau_0^{-1}(\frac{\kappa v q_F^3}{\mu^2})^2}$\\
 		$\tau_{z,e1}^{-1}=\pi\nu (A_{xz}^2+A_{yz}^2)\frac{v^2q_F^2}{\mu^2}$  & $\tau_{z,e2}^{-1}=\pi\nu A_{zz}^2(1+\frac{E_g^2}{\mu^2})$  
 		& $\tau_{z,e}^{-1}=\tau_{z,e1}^{-1}+\tau_{z,e2}^{-1}$  & $\textcolor{black}{\tau_{z,e1}^{-1},\tau_{z,e2}^{-1}\propto \tau_0^{-1}(\frac{\Delta_{KM}v^2q_F^2}{\mu^3})^2}$\\
 		$\tau_{z,o1}^{-1}=\pi\nu \sum_{i,j=x,y}(A_{ij}^2)\frac{v^2q_F^2}{\mu^2}\quad\,$& $\tau_{z,o2}^{-1}=\pi\nu (A_{zx}^2+A_{zy}^2)(1+\frac{E_g^2}{\mu^2})$ &$ \tau_{z,o}^{-1}=\tau_{z,o1}^{-1}+\tau_{z,o2}^{-1}$ & $\textcolor{black}{\tau_{z,o1}^{-1}, \tau_{z,o2}^{-1} \propto \tau_0^{-1}(\frac{\lambda v q_F }{\mu^2})^2}$\\
 		$\tau_{zv,e}^{-1}=\pi\nu A_{0z}^2(1+\frac{E_g^2}{\mu^2})$ &&& $\textcolor{black}{\tau_{zv,e}^{-1}\propto \tau_0^{-1}(\frac{\kappa v \Delta_{KM}q_F^3}{\mu^3})^2}$\\
 		$\tau_{zv,o}^{-1}=\pi \nu (A_{0x}^2+A_{0y}^2)(1+\frac{E_g^2}{\mu^2})$ &&& $\textcolor{black}{\tau_{zv,o}^{-1}\propto \tau_0^{-1}(\frac{\lambda E_gv q_F}{\mu^3})^2}$\\
 		\hline
 		\textcolor{black}{Intervalley scattering rates} & & & \textcolor{black}{Estimates} \\
 		\hline
 		\multicolumn{3}{l|}{$\tau_{iv}^{-1}=\pi\nu\sum_{i=x,y}[\textcolor{black}{2\sum_{j=\pm}V_{ji}(1+j\frac{E_g}{\mu})^2} + (V_{xi}^2)\frac{v^2q_F^2}{\mu^2}] $} &  $\qquad$ / \\
 		$\tau_{iv,e}^{-1}=\pi \nu (M_{zx}^2+M_{zy}^2) \frac{v^2q_F^2}{\mu^2}$ &&& $\textcolor{black}{\tau_{iv,e}^{-1}\propto \textcolor{black}{\tau_{iv}^{-1}(\frac{ \Delta_{KM}v q_F}{\mu^2})^2}}$\\
 		$\tau_{iv,o}^{-1}=\pi \nu \sum_{i,j=x,y} (M_{ij}^2) \frac{v^2q_F^2}{\mu^2}$ &&& $\textcolor{black}{\tau_{iv,o}^{-1}\propto \tau_{iv}^{-1}(\frac{\lambda v q_F}{\mu^2})^2}$\\
 		\hline
 		\hline
 	\end{tabular}
 	\caption{\label{table1} Left: Dominant diagonal scattering rate, $\tau_0^{-1}$, and \textcolor{black}{the 11 other} independent scattering rates \cite{foot1} originating from the disorder Hamiltonian \eqref{eqn3}. The notation for the scattering rates was taken and adapted from Ref.~\onlinecite{mccann2012z}. The index $z$ indicates that the related disorder potential is sublattice dependent. $zv$ and $iv$ indicate coupling to the valley matrices $\eta_z$ and $\eta_{x,y}$, respectively. Indices $e$ and $o$ indicate coupling to the spin matrices $s_z$ and $s_{x,y}$, respectively. Spin-independent disorder is represented by the rates $\tau_0^{-1}, \tau_z^{-1}$\textcolor{black}{,} and $\tau_{iv}^{-1}$, which describe diagonal, intervalley, and sublattice-dependent intravalley disorder. Spin-dependent disorder is represented by the rates $\tau_{i,j}^{-1}$ ($i=z,zv,iv; j=e,o$), which describe intra- ($i=z,zv$) or intervalley ($i=iv$), and spin-preserving ($j=e$) or spin-flipping ($j=o$) disorder. Right: Estimates of the \textcolor{black}{phenomenological scattering rates}, \textcolor{black}{obtained by the combination of band structure parameters and potential disorder only,} \textcolor{black}{assuming that all intervalley components of the potential disorder are of similar strength.} }
 \end{table*}

 \textcolor{black}{Assuming that only potential disorder is present in the system, we can estimate the phenomenological scattering rates, related with the parameters in Eq.~\eqref{eqn4}, as shown in the right column of Table \ref{table1}. We do so by comparing the disorder terms generated by $H^D_{\mathbf{qq'}}$ after rotation and projection onto the conduction band with the terms generated by $H^{D0}_{\mathbf{qq'}}$ only, but taking into account corrections up to order $1/\mu$. In this way, we can relate the phenomenological disorder parameters with the main Hamiltonian~\eqref{eqn1} and the magnitude of the potential disorder. }
 
\begin{figure}[h!]
	\includegraphics[width=0.4\textwidth]{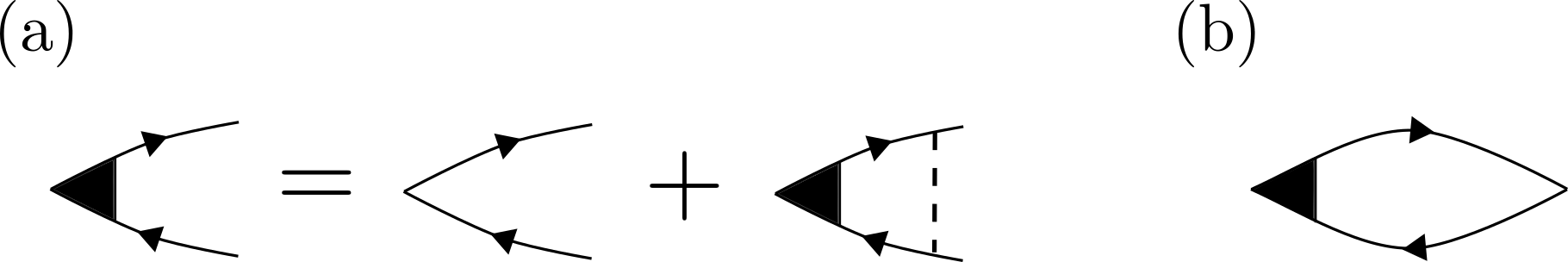}
	\caption{\label{figure1} (a) Vertex renormalization. (b) Drude conductivity diagram. Solid arrows represent Green's functions, while the dashed lines represent disorder. The upper (lower) branch of the diagrams corresponds to retarded (advanced) Green's functions.  }
\end{figure}

The current operator in the projected basis is given by $\mathcal{J}_{x\mathbf{q}}=v_F\cos \theta$. Due to the anisotropy of the disorder potential, the current vertex is renormalized, as illustrated in diagrammatic form in Fig.~\ref{figure1}(a). \textcolor{black}{Namely}, the bare vertex is dressed by a series of ladder diagrams, known as diffusons. The renormalized vertex is then given as
\begin{equation}
\tilde{\mathcal{J}}_{x\mathbf{q}}=\frac{\tau_{tr}}{\tau_0}\mathcal{J}_{x\mathbf{q}} \quad \text{with} \quad \tau_{tr}=
\bigg(1+\frac{v^2q_F^2}{4E_g^2+v^2q_F^2}\bigg)\tau_0.
\label{eqn8}
\end{equation}
Here, we have introduced the transport time $\tau_{tr}$, which takes the value $\tau_0$ at the bottom of the conduction band $\mu\approx E_g$, where the spectrum is parabolic (similarly to conventional metals), and $2\tau_{0}$ deep in the conduction band $\mu\gg E_g$, where the spectrum is linear (as in graphene)\cite{mccann2006weak}. The Drude conductivity is then given as
\begin{equation}
\sigma=\frac{e^2}{2\pi}\int \frac{d^2\mathbf{p}}{(2\pi)^2}\text{Tr}\bigg[\tilde{\mathcal{J}}_{x\mathbf{q}}G^R_{\mathbf{q}\omega}\mathcal{J}_{x\mathbf{q}}G^A_{\mathbf{q}\omega}\bigg]\Bigg|_{\omega=0}=4e^2\nu D,
\label{eqn9}
\end{equation}
where $D=\frac{1}{2}v_F^2\tau_{tr}$ is the diffusion constant, and the factor $4$ originates from spin and valley degeneracy. The corresponding diagram is shown in Fig.~\ref{figure1}(b).

\section{Interference correction}\label{sec3}
\begin{figure}[h!]
	\includegraphics[width=0.40\textwidth]{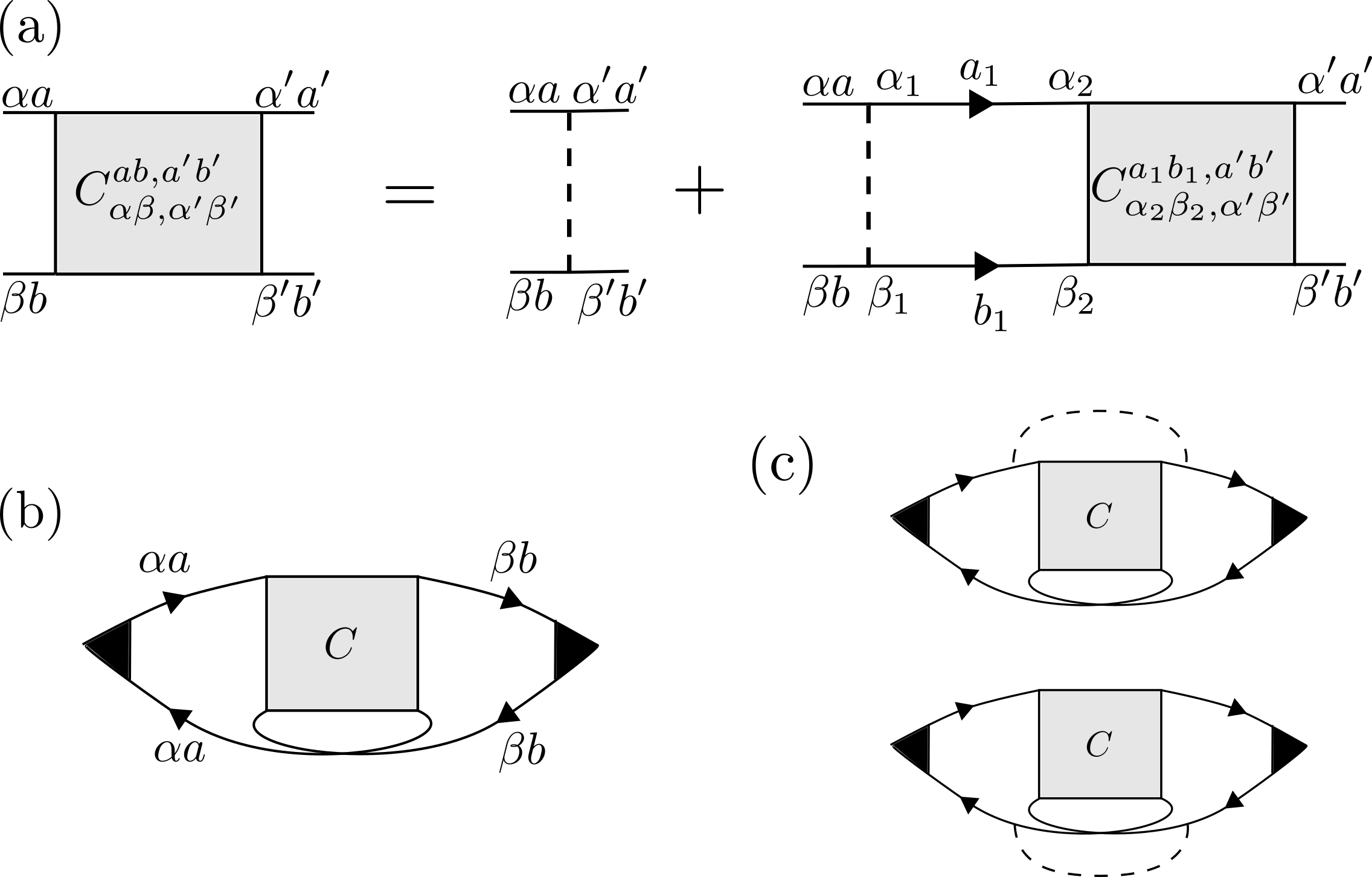}
	\caption{(a) Bethe-Salpeter equation for the Cooperons. (b) Bare Hikami box. The Hikami boxes with external lines that are diagonal in spin-space give a dominant contribution to the quantum correction in the diffusive limit. (c) Dressed Hikami boxes. For the definition of diagram elements, see Fig.~\ref{figure1}. Greek indices in the subscript describe spin, while Latin indices in the super\textcolor{black}{s}cript describe the valley degree of freedom.}
	\label{figure2}
\end{figure}

The interference correction to the Drude conductivity~\eqref{eqn9} can be expressed in terms of Cooperons, $C_{\alpha\beta,\alpha'\beta'}^{ab,a'b'}$, which represent disorder averages of two Green's functions and correspond to maximally crossed diagrams \cite{akkermans2007mesoscopic}.
 The Greek indices in the subscript (Latin indices in the superscript) correspond to the spin (valley) degree of freedom and take values $\pm 1$. 
The Cooperons are determined from a system of coupled Bethe-Salpeter equations, as shown in diagrammatic form in Fig.~\ref{figure2}(a). Namely, 
\begin{widetext}
\begin{equation}
C_{\alpha\beta,\alpha'\beta'}^{ab,a'b'}(\theta,\theta';\mathbf{Q})=W_{\alpha\beta,\alpha'\beta'}^{ab,a'b'}(\theta,\theta')+\int_{0}^{2\pi}\frac{d\theta''}{2\pi}
W_{\alpha\alpha_1,\beta\beta_1}^{a a_1,b b_1}(\theta,\theta'')\Pi_{\alpha_1\beta_1,\alpha_2\beta_2}^{a_1b_1}(\theta'';\mathbf{Q})
C_{\alpha_2\beta_2,\alpha'\beta'}^{a_1b_1,a'b'}(\theta'',\theta';\mathbf{Q}).
\label{eqn10}
\end{equation}		
Here, summation over repeated indices is assumed, and we have introduced the disorder correlator $W$ and the polarization operator $\Pi$ as
\begin{equation}
W_{\alpha\beta,\alpha'\beta'}^{ab,a'b'}(\theta,\theta')=\langle[\mathcal{H}^{D}_{\mathbf{qq'}}]_{\alpha\alpha'}^{aa'} [\mathcal{H}^{D}_{\mathbf{\bar{q}\bar{q}'}}]_{\beta\beta'}^{b b'}\rangle \qquad \text{and} \qquad 
\Pi_{\alpha\beta,\alpha'\beta'}^{ab}(\theta; \mathbf{Q})=
\nu\int d\xi_{\mathbf{q}}[G^R_{\mathbf{q}\epsilon+\omega}]_{\alpha\alpha'}^{a}[G^A_{\mathbf{\bar{q}+Q}\omega}]_{\beta\beta'}^{b},
\label{eqn11}
\end{equation}
\textcolor{black}{respectively}. Note that the Green's functions are diagonal in valley space, so the polarization operator only depends on two valley indices. The weak localization correction $\delta\sigma$ can now be expressed in terms of Cooperons as
 \begin{equation}
 \delta\sigma=\frac{e^2}{2\pi}\int\frac{d^2\mathbf{Q}}{(2\pi)^2}\int_0^{2\pi}\frac{d\,\theta}{2\pi}\frac{d\,\theta'}{2\pi}
 4\pi\nu\tau_0^3 \bigg[2\pi \delta(\theta-\theta') -2\pi \nu\tau_0 W_{\alpha\beta,\alpha\beta}^{ab,ab}(\theta,\theta')\bigg]\tilde{\mathcal{J}}_{x\mathbf{q}}\tilde{\mathcal{J}}_{x\bar{\mathbf{q}}'}C^{ab,ba}_{\alpha\beta,\beta\alpha}(\theta,\bar{\theta}';\mathbf{Q}).
 \label{eqn12}
\end{equation}
\end{widetext}
Here, the first contribution in the square bracket comes from the bare Hikami box \cite{akkermans2007mesoscopic} [shown in Fig.~\ref{figure2}(b)], while the second one comes from two Hikami boxes dressed by an intravalley impurity line [shown in Fig.~\ref{figure2}(c)].

\textcolor{black}{We proceed by solving Eq.~\eqref{eqn10} in the presence of the dominant diagonal scattering only, in Sec.~\ref{subsec1}. Next, we include all other types of disorder in Sec.~\ref{subsec2}. Finally, the interference-induced magnetoconductance and the main result of our work are presented in Sec.~\ref{subsec3}.}

\textcolor{black}{\subsection{Cooperons in the presence of diagonal disorder only \label{subsec1}}}
\textcolor{black}{In order to resolve the angular structure of the Cooperons, we will first consider the case where only the diagonal disorder with rate $\tau_0^{-1}$ is present. The other types of scattering will not affect this structure, but only introduce additional Cooperon gaps. Furthermore, the angular structure is independent of the spin structure. Therefore, we also neglect the spin structure here, setting $\Delta_{so}$ and $h$ to zero. To simplify the notation, spin indices are omitted in this subsection.} 
 
 We proceed with this calculation in the same spirit as in  Ref.~\onlinecite{shan2012spin}. First, we expand the Cooperons and the disorder correlator in harmonics,
 \begin{align}
 C^{ab,a'b'}(\theta,\theta';\mathbf{Q})&=\,\sum_{\textcolor{black}{n,m}=-\infty}^{\infty}C^{ab,a'b'}_{\textcolor{black}{nm}}(\mathbf{Q})e^{-i(\textcolor{black}{n}\theta-\textcolor{black}{m}\theta')},   \nonumber \\ W^{ab,a'b'}(\theta,\theta')&=\,\sum_{n=-\infty}^{\infty}W^{ab,a'b'}_n e^{-in(\theta-\theta')}.
 \label{eqn13}
 \end{align}
Furthermore, $a=a'$ and $b=b'$ in the absence of intervalley scattering. The only Cooperon that enters the interference correction \eqref{eqn12} is the intravalley one, $C^{aa,aa}(\theta,\theta')$. From Eqs.~\eqref{eqn10} and ~\eqref{eqn13}, we get a system of coupled equations for its harmonics, whose solution yields
 \begin{align}
 &C^{aa,aa}(\theta,\theta';\mathbf{Q})= C_{\textcolor{black}{00}}^{aa,aa}(\mathbf{Q})+C_{\textcolor{black}{aa}}^{aa,aa}(\mathbf{Q})e^{-ia(\theta-\theta')} 
 \nonumber \\
 &\text{with} \quad
 C_{\textcolor{black}{ii}}^{aa,aa}(\mathbf{Q})= \frac{1}{2\pi\nu\tau_0^2}\frac{1}{D_i |\mathbf{Q}|^2-i\omega+\tau_\phi^{-1}+\Gamma_i}.
 \label{eqn14}
 \end{align}
Here, $a=\pm 1$, $\Gamma_0=\frac{1}{\tau_0}\frac{(\mu-E_g)^2}{(\mu+E_g)^2}$ and $\Gamma_a=\frac{1}{\tau_0}\frac{2E_g^2}{\mu^2-E_g^2}$ are the relevant Cooperon gaps, and $D_0=\frac{1}{8}v_F^2\tau_0(3+\frac{E_g^2}{\mu^2})$ and $D_a=v_F^2\tau_0 \frac{(E_g^2+\mu^2)^2}{(\mu^2-E_g^2)^2}$ are diffusion constants. Furthermore, we introduced the inelastic dephasing rate, $\tau_\phi^{-1}$.

 We see that, in general, both $C_{\textcolor{black}{00}}$ and $C_{\textcolor{black}{aa}}$ will have a large gap of the order $\tau_0^{-1}$ and, thus,  will be suppressed in the diffusive limit, except in two special cases. Firstly, $\Gamma_0$ vanishes at $\mu=E_g$. Close to the band bottom, for $\mu/E_g-1\lesssim 2\sqrt{\tau_0/\tau_\phi}$, one finds $\Gamma_0\lesssim\tau_\phi^{-1}$. Thus, in this regime, the Cooperon $C_{\textcolor{black}{00}}$ is not suppressed. Secondly, $\Gamma_a$ vanishes for $\mu\rightarrow\infty$. Thus, deep in the band, at $\mu/E_g \gtrsim \sqrt{2\tau_\phi/\tau_0}$, one finds $\Gamma_a\lesssim\tau_\phi^{-1}$, and the Cooperon $C_{\textcolor{black}{aa}}$ is not suppressed \textcolor{black}{either}. Higher-order harmonics, although non-zero, will always have a non-vanishing gap of the order $\tau_0^{-1}$ and will be neglected. We can therefore write 
\begin{align}
&C^{aa,aa}(\theta,\bar{\theta};\mathbf{Q})=\frac{\Xi}{2\pi\nu \tau_0^2}\frac{1}{D|\mathbf{Q}|^2-i\omega+\tau_\phi^{-1}+\Gamma_\Xi}, \nonumber \\
&\text{where }\Xi=\begin{cases}
1,  &\frac{\mu}{E_g}-1 \lesssim 2\sqrt{\frac{\tau_0}{\tau_\phi}},  \\
0,  &2\sqrt{\frac{\tau_0}{\tau_\phi}} \ll\frac{\mu}{E_g}-1\ll \sqrt{\frac{2\tau_\phi}{\tau_0}}, \\
-1,   &\frac{\mu}{E_g} \gtrsim \sqrt{\frac{2\tau_\phi}{\tau_0}},
\end{cases}
\label{eqn15}
\end{align}
and $\Gamma_1=\tau_0^{-1}[vq_F/(2\mu)]^4$, $\Gamma_{-1}=2\tau_0^{-1}(E_g/\mu)^2$. Note that the diffusion constants $D_0$ and $D_a$ reduce to $D$, introduced in Eq.~\eqref{eqn9}, in the relevant limits. 

Upon inserting Eq.~\eqref{eqn15} into Eq.~\eqref{eqn12}, we obtain the quantum correction for massive Dirac fermion systems in the presence of smooth disorder, consistent with Ref.~\onlinecite{shan2012spin}. Its behavior is governed by the doping-dependent coefficient $\Xi$: for a \textcolor{black}{large} Dirac mass \textcolor{black}{$E_g$} ($\Xi=1$)\textcolor{black}{,} we get WL, whereas in the massless system ($\Xi=-1$), we get WAL. The quantum correction vanishes in the intermediate mass regime. This can be reinterpreted \cite{shan2012spin} in terms of the Berry phase of a massive Dirac material given as $\varphi_B=\pi (1-E_g/\mu)$, which introduces no phase shift to the electron interference in the large mass limit (leading to WL), and a shift of $\pi$ for massless systems (leading to WAL). 

Next, we will find the intervalley Cooperon $C^{a\bar{a},a \bar{a}}(\theta,\theta')$. Note that it does not enter the quantum correction \eqref{eqn12}, but it is useful to resolve its angular structure for later use.  We find that the only harmonic that is not gapped is $C_{\textcolor{black}{00}}$, and we can write
\begin{align}
C^{a\bar{a},a\bar{a}}(\theta,\theta';\mathbf{Q})&=\, C_{\textcolor{black}{00}}^{a\bar{a},a\bar{a}}(\mathbf{Q})  \nonumber \\
&=\frac{1}{2\pi\nu \tau_0^2} \frac{1}{D|\mathbf{Q}|^2-i\omega+\tau_\phi^{-1}}.
\label{eqn16}
\end{align}

\textcolor{black}{\subsection{Cooperons in the presence of all disorder terms \label{subsec2}}}
 We proceed to solve the Cooperon equation \eqref{eqn10} in the presence of all disorder terms. Additional intervalley Cooperons of the form $C^{a\bar{a},\bar{a}a}$ can now exist. Since they are coupled to $C^{a\bar{a},a\bar{a}}$ via intervalley scattering, which does not introduce additional angular dependence, they will also be angularly-independent. Using Eqs.~\eqref{eqn15} and \eqref{eqn16}, we can write for all Cooperons
 \begin{align}
 C^{ab,a'b'}(\Xi; \mathbf{Q})=& \,[C_{\textcolor{black}{00}}^{aa,aa}(\mathbf{Q})\delta_{\Xi,1}+C_{\textcolor{black}{aa}}^{aa,aa}(\mathbf{Q})\delta_{\Xi,-1}]\nonumber \\ 
 &\times \delta_{aa'}\delta_{bb'}\delta_{ab}+ 
 C_{\textcolor{black}{00}}^{a\bar{a},b\bar{b}}(\mathbf{Q})\delta_{a\bar{b}}\delta_{a'\bar{b}'}\textcolor{black}{,} \nonumber  \\
 W^{ab,a'b'}(\Xi)=& \,[W_0^{aa,aa}\delta_{\Xi,1}+W_a^{aa,aa}\delta_{\Xi,-1}] \nonumber \\
& \times \delta_{aa'}\delta_{bb'}\delta_{ab}+W_0^{a\bar{a},b\bar{b}}\delta_{a\bar{b}}\delta_{a'\bar{b}'}.
\label{eqn17}
 \end{align}
 Then,  Eq.~\eqref{eqn10} can be written in a simpler, angularly-independent form,
 \begin{align}
& C_{\alpha\beta,\alpha'\beta'}^{ab,a'b'}(\Xi;\mathbf{Q})=W_{\alpha\beta,\alpha'\beta'}^{ab,a'b'}(\Xi)\nonumber \\&+ 
W_{\alpha\alpha_1,\beta\beta_1}^{a a_1,b b_1}(\Xi)\Pi_{\alpha_1\beta_1,\alpha_2\beta_2}^{a_1b_1}(\mathbf{Q})
 C_{\alpha_2\beta_2,\alpha'\beta'}^{a_1b_1,a'b'}(\Xi;\mathbf{Q}).
 \label{eqn18}
 \end{align}
 Next, we employ a transformation to the singlet-triplet basis \cite{mccann2012z} in spin and valley space,
 \begin{equation}
 \textcolor{black}{M}_{ss'}^{ll'}=\frac{1}{4} [s_ys_s]_{\alpha\beta}[\eta_x\eta_l]^{ab}\textcolor{black}{M}^{ab,a'b'}_{\alpha\beta,\alpha'\beta'}
 [s_{s'}s_y]_{\beta'\alpha'}[\eta_{l'}\eta_x]^{b'a'},
 \label{eqn19}
 \end{equation}
 where indices $s,s'=0$ and $l,l'=0$ correspond to spin- and valley-singlet Cooperon modes, respectively, while $s,s'=x,y,z$ and $l,l'=x,y,z$  correspond to spin- and valley-triplet modes. Here, the operator $\textcolor{black}{M}$ can stand for a Cooperon ($C$), disorder correlator ($W$), or a polarization operator ($\Pi$). The disorder correlator is diagonal in the singlet-triplet space, $W_{ss'}^{ll'}(\Xi)=W_s^l(\Xi)\delta_{ss'}\delta_{ll'}$, and the Cooperon equation \eqref{eqn18} after the transformation becomes
 \begin{equation}
 C_{ss'}^{ll'}(\Xi;\mathbf{Q})=W_{s}^{l}(\Xi)\delta_{ss'}\delta_{ll'}+ W_{s}^{l}(\Xi)\Pi_{ss_1}^{ll_1}(\mathbf{Q})C_{s_1s'}^{l_1l'}(\Xi;\mathbf{Q}).
 \label{eqn20}
 \end{equation}
 The quantum correction involves only the diagonal Cooperons $C_{ss}^{ll}\equiv C_s^l$. Note that triplets modes $C_{\textcolor{black}{s}}^x$ and $C_{\textcolor{black}{s}}^y$ are related to the intravalley Cooperons, while the valley-singlet $C_{\textcolor{black}{s}}^0$ and triplet $C_{\textcolor{black}{s}}^z$ are related to intervalley ones. \textcolor{black}{Finally, the interference correction, Eq.~\eqref{eqn12}, in the new basis has the form}
 \begin{align}
 \delta\sigma&=-\frac{e^2 D}{\pi}(2\pi\nu\tau_0^2)\int\frac{d^2\mathbf{Q}}{(2\pi)^2}\times  \nonumber \\
  & \sum_s c_s\bigg[ \sum_{l=0,z} c^l C_s^l(\Xi; \mathbf{Q})+
 \Xi \sum_{l=x,y}c^l C_s^l(\Xi;\mathbf{Q})
 \bigg],
 \label{eqn21}
 \end{align}
 where $c_s=-1,1,1,1$ and $c^l=1,1,1,-1$ for $s,l=0,x,y,z$. Eq.~\eqref{eqn21} generalizes similar expressions from Refs.~\onlinecite{mccann2012z} and \onlinecite{ochoa2014spin}, which are valid at $\Xi=-1$ and $\Xi=1$, respectively.
 
 The diagonal Cooperon modes $C_s^l$, necessary to compute $\delta\sigma$, are determined by solving Eq.~\eqref{eqn20}. Due to the spin-splitting described by $\Delta_{so}$ and $h$, the polarization operator $\Pi_{ss'}^{ll'}(\mathbf{Q})$ is not diagonal in the singlet-triplet space. As a consequence, some Cooperon modes are coupled. As will be discussed in the further text, \textcolor{black}{the coupling of different Cooperon modes by the spin-splitting fields suppresses them. In a physical sense, Cooperons coupled by the fields describe interference of electrons coming from two spin-split bands, which is suppressed by the energy difference of the electrons. On the other hand, interference of electrons in degenerate bands is described by the non-coupled Cooperons. Note that momentum dependent parts of the Hamiltonian \eqref{eqn1}, such as Rashba SOC and trigonal warping, do not cause coupling of different Cooperon modes in the diffusive limit, but only enter their gaps.} 
 
\paragraph*{\textcolor{black}{a. \, Non-coupled Cooperon modes.}} First, we solve the Cooperons that are not coupled by the valley-dependent SOC or the in-plane field, with the indices   $(s,l)=(y,x),(y,y),(z,0),(z,z)$. They are given by
 \begin{equation}
 C_{s}^{l}=\frac{1}{2\pi\nu \tau_0^2}\frac{1}{\mathcal{P}_s^{l}}.
 \label{eqn22}
 \end{equation}
 Here, we introduced $\mathcal{P}_s^{l}=D|\mathbf{Q}|^2-i\omega+\tau_{\phi}^{-1}+\Gamma_{s}^{l}$, where the Cooperon gaps $\Gamma_s^l$ are specified in Table \ref{table2}.  Because the intravalley Cooperons have different angular dependence in the two extreme limits of Eq.~\eqref{eqn15}, their gaps $\Gamma_s^x$ and $\Gamma_s^y$ will also depend on the relevant limit (right-hand side of Table \ref{table2}). Intervalley Cooperons, on the other hand, do not depend on angles and chemical potential and have the same gaps for any $\mu$ (left-hand side of Table \ref{table2}).
 
 The Cooperon gaps contain the scattering rates originating  from the phenomenological disorder potential \eqref{eqn3}. Their estimates, listed in Table \ref{table1}, are inversely proportional to the scattering times $\tau_0$ and $\tau_{iv}$. These rates are therefore induced and reinforced by disorder, and behave similarly to the Elliott-Yafet spin relaxation mechanism  \cite{elliott1954theory, yafet1963g}. \textcolor{black}{This includes the well-known scattering rate due to the Kane-Mele SOC \cite{mccann2012z}, captured by $\tau_{z,e}^{-1}\propto \tau_0^{-1}(\frac{\Delta_{KM}v^2q_F^2}{\mu^3})^2$ (see Table \ref{table1}}). Additionally, scattering rates that are proportional to the potential scattering time $\tau_0$  also enter the gaps:
 \begin{equation}
 \textcolor{black}{\tau_{BR}^{-1}=2\bigg(\frac{\lambda vq_F}{\mu}\bigg)^2\tau_{tr}, \quad 
 \tau_{W}^{-1}=2 \bigg( \frac{\kappa v q_F^3}{\mu}\bigg)^2\tau_0.}
 \label{eqn23}
 \end{equation}
 They are related with  Rashba SOC and trigonal warping, respectively. These rates appear since electrons, due to the details of the band structure, acquire an additional phase upon propagation in-between two scattering events. This effect is suppressed by disorder. The first rate in Eq.~\eqref{eqn23} is associated with the Dyakonov-Perel \cite{d1971spin} spin relaxation mechanism. The second rate describes the suppression of intravalley Cooperons due to the breaking of rotational symmetry by trigonal warping, as discussed in Ref.~\onlinecite{mccann2006weak}.

  \paragraph*{\textcolor{black}{b. \, Coupled Cooperon modes}.} Next, we address the coupled Cooperon modes. The effect of the in-plane Zeeman field $h$ applied along the $x$-direction is such that it couples the spin-singlet $C_0^{\textcolor{black}{l}}$ and spin-triplet $C_x^{\textcolor{black}{l}}$ \textcolor{black}{Cooperons}, as discussed for conventional metals \cite{maekawa1981magnetores}. Valley-dependent SOC behaves similarly to an effective Zeeman field in $z$-direction, but acts differently from the true Zeeman field as it does not break the time-reversal symmetry, and therefore does not affect the spin- and valley-singlet $C_0^0$, which is protected by this symmetry. It couples the Cooperons $C_{x}^{0(z)}$ with $C_{y}^{z(0)}$, and $C_{0}^{x(y)}$ with $C_{z}^{y(x)}$, as discussed in Ref.~\onlinecite{ochoa2014spin}. The equations for all the coupled Cooperon modes can be compactly written in a matrix form
  \begin{widetext}
 \begin{align}
  \begin{bmatrix}
  \mathcal{P}_0^{x(y)} & \mp 2\Delta_{so} & -2ih \\
  \pm 2\Delta_{so} & \mathcal{P}_z^{y(x)} & 0\\
  -2ih & 0 & \mathcal{P}_x^{x(y)}
  \end{bmatrix}
  \begin{bmatrix}
  C_{00}^{xx(yy)} & C_{0z}^{xy(yx)} & C_{0x}^{xx(yy)}\\
  C_{z0}^{yx(xy)} & C_{zz}^{yy(xx)} & C_{zx}^{yx(xy)} \\
  C_{x0}^{xx(yy)} & C_{xz}^{xy(yx)} & C_{xx}^{xx(yy)}
  \end{bmatrix}= \frac{1}{2\pi\nu\tau_0^2}=
 \begin{bmatrix}
 \mathcal{P}_x^{0(z)} & -2\Delta_{so} & -2 i h \\
 2\Delta_{so} & \mathcal{P}_y^{z(0)} & 0 \\
 -2 i h & 0 & \mathcal{P}_0^{0(z)}
 \end{bmatrix}
 \begin{bmatrix}
 C_{xx}^{00(zz)} & C_{xy}^{0z(z0)} & C_{x0}^{00(zz)}\\
 C_{yx}^{z0(0z)} & C_{yy}^{zz(00)} & C_{y0}^{z0(0z)}\\
 C_{0x}^{00(zz)} & C_{0y}^{0z(z0)} & C_{00}^{00(zz)}
 \end{bmatrix}.
  \label{eqn24}
 \end{align}
 \end{widetext}

Eq.~\eqref{eqn24} summarizes 4 matrix equations, each involving 3 coupled modes. Since the Green's functions are diagonal in valley space, the equations for intra- and intervalley Cooperons are decoupled. This can be seen in Eq.~\eqref{eqn24}, where the left-hand (right-hand) side describes matrix equations for intravalley (intervalley) Cooperon modes. 
\begin{table*}[t]
	\begin{tabular}{l| l}
		\hline \hline
		Relaxation gaps for $C^0$ and $C^z$ & Relaxation gaps for $C^x$ and $C^y$ at $\Xi=-1$  \\
		\hline
		$\Gamma_0^0=0$ & 	$\Gamma_x^x=\Gamma_y^x=\Gamma_x^y=\Gamma_y^y=\tau_*^{-1}+2\tau_{z,e}^{-1}+\textcolor{black}{\tau_{z,o}^{-1}}+\tau_{zv,o}^{-1}+\textcolor{black}{\tau_{BR}^{-1}}$ \\
		$\Gamma_x^0=\Gamma_y^0=2\tau_{z,e}^{-1}+\textcolor{black}{\tau_{z,o}^{-1}}+2\textcolor{black}{\tau_{zv,e}^{-1}}+\tau_{zv,o}^{-1}+2\tau_{iv,e}^{-1}+\tau_{iv,o}^{-1}+\textcolor{black}{\tau_{BR}^{-1}}\textcolor{black}{=\tau_s^{-1}}$ & $\Gamma_0^x=\Gamma_0^y=\tau_*^{-1}+2\textcolor{black}{\tau_{zv,e}^{-1}}+2\tau_{zv,o}^{-1}$\\ 
		$\Gamma_z^0=\textcolor{black}{2\tau_{z,o}^{-1}}+2\tau_{zv,o}^{-1}+2\tau_{iv,o}^{-1}+\textcolor{black}{2\tau_{BR}^{-1}}\textcolor{black}{=2\tau_{asy}^{-1}}$ & $\Gamma_z^x=\Gamma_z^y=\tau_*^{-1}+\textcolor{black}{2\tau_{z,o}^{-1}}+2\textcolor{black}{\tau_{zv,e}}^{-1}+2\textcolor{black}{\tau_{BR}^{-1}}$  \\ \cline{2-2}
		$\Gamma_0^z=2\tau_{iv}^{-1}+2\tau_{iv,e}^{-1}+2\tau_{iv,o}^{-1}$ & Relaxation gaps for $C^x$ and $C^y$ at $\Xi=1$\\ \cline{2-2}
		$\Gamma_x^z=\Gamma_y^{z}=2\tau_{iv}^{-1}+2\tau_{z,e}^{-1}+\textcolor{black}{\tau_{z,o}^{-1}}+2\textcolor{black}{\tau_{zv,e}^{-1}}+\tau_{zv,o}^{-1}+\tau_{iv,o}^{-1}+\textcolor{black}{\tau_{BR}^{-1}}$ & $\Gamma_x^x=\Gamma_y^x=\Gamma_x^y=\Gamma_y^y=\tau_{**}^{-1}+\tau_{z,o2}^{-1}+\tau_{zv,o}^{-1}+\tau_{BR}^{-1}$\\
		$\Gamma_z^z=2\tau_{iv}^{-1}+\textcolor{black}{2\tau_{z,o}^{-1}}+2\tau_{zv,o}^{-1}+2\tau_{iv,e}^{-1}+\textcolor{black}{2\tau_{BR}^{-1}}$ & $\Gamma_0^x=\Gamma_0^y=\tau_{**}^{-1}+2\tau_{z,e2}^{-1}+2\tau_{z,o2}^{-1}+2\tau_{zv,e}^{-1}+2\tau_{zv,o}^{-1}$\\
		& $\Gamma_z^x=\Gamma_z^y=\tau_{**}^{-1}+2\tau_{z,e2}^{-1}+2\textcolor{black}{\tau_{zv,e}^{-1}}+2\tau_{BR}^{-1}$\\
		\hline
		\multicolumn{2}{l}{ $\tau_*^{-1}=\tau_{iv}^{-1}+2\tau_z^{-1}+\tau_{iv,e}^{-1}+\tau_{iv,o}^{-1}+\tau_{W}^{-1}+\frac{2}{\tau_0}\frac{E_g^2}{\mu^2}$ } \\
		\multicolumn{2}{l}{
		$ \tau_{**}^{-1}=\tau_{iv}^{-1}+\tau_{z1}^{-1}+\tau_{z,e1}^{-1}+\tau_{z,o1}^{-1}+\tau_{iv,e}^{-1}+\tau_{iv,o}^{-1}+\tau_{W}^{-1}+\frac{1}{16\tau_0}\frac{v^4q_F^4}{\mu^4}$} \\
		\hline\hline
	\end{tabular}
	\caption{\label{table2} Left: Relaxation gaps $\Gamma_s^l$ for intervalley Cooperons, where indices $s$ and $l$ denote spin and valley, respectively. There are 8 intervalley Cooperons. The time-reversal symmetry sets the gap $\Gamma_0^0$ to zero, while the $x-y$ symmetry imposes equality of all $x$ and $y$ spin-triplet gaps. As a result, there are only 5 independent gaps. \textcolor{black}{The scattering rates $\tau_{asy}^{-1}$ and $\tau_s^{-1}=\tau_{sym}^{-1}+\tau_{asy}^{-1}$, related to the valley-singlet gaps $\Gamma_{i}^0$ ($i=x,y,z$), are introduced in Eqs.~\eqref{eqn30} and \eqref{eqn31}.} Right: Relaxation rates for intravalley Cooperons, which depend on the \textcolor{black}{chemical potential}, captured by the coefficient $\Xi$. \textcolor{black}{In each regime}, there are \textcolor{black}{8} intravalley Cooperons.  $x-y$ symmetry imposes equality of all $x$ and $y$ triplet gaps, in both spin and valley space. As a result, there are only \textcolor{black}{3} independent gaps. Since at $\Xi=0$ intravalley Cooperons do not contribute to the quantum correction, the related gaps are not included in the table. For a definition of the different scattering rates, see Table~\ref{table1}.  }
\end{table*}

 \textcolor{black}{Note that in, Eq.~\eqref{eqn24}, the spin-splitting fields, $h$ and $\Delta_{so}$, are considered only up to the leading order in $\tau_0$ in the diffusive limit. As discussed in Appendix~\ref{ap1}, by considering higher-order terms, we find that these fields also modify the Cooperon gaps by supplementing them with terms of the order $\Delta_{so}^2\tau_0$ and $h^2\tau_0$. However, these terms can always be neglected, as their effect is small compared to the one produced by coupling of the Cooperon modes by these fields.}
\textcolor{black}{\subsection{Interference-induced magnetoconductance \label{subsec3}}}
	Finally, after inverting the matrices in Eq.~\eqref{eqn24}, we obtain all Cooperon modes. Combining them with Eq.~\eqref{eqn21}, and introducing the conductance quantum $\sigma_0=e^2/(2\pi^2 \hbar)$,  we arrive at the expression for the interference correction 
\begin{widetext}
\textcolor{black}{\begin{align}
\delta\sigma= & \;2\pi\sigma_0 D\int \frac{d^2\mathbf{Q}}{(2\pi)^2}\bigg[
-\Xi\bigg(\frac{1}{\mathcal{P}_y^x}+\frac{1}{\mathcal{P}_y^y}+\mathcal{A}(_{z}^{y},_{x}^{x},_{0}^{x}) +\mathcal{A}(_{z}^{x},_{x}^{y},_{0}^{y})\bigg) 
-\frac{1}{\mathcal{P}_z^0}+\frac{1}{\mathcal{P}_z^z}+\mathcal{A}(_{y}^{z},_{0}^{0},_{x}^{0})-\mathcal{A}(_{y}^{0},_{0}^{z},_{x}^{z})
\bigg],\,\nonumber \\
&\text{ where}  \quad \mathcal{A}(_{s_1}^{l_1},_{s_2}^{l_2},_{s_3}^{l_3})=
2\pi\nu\tau_0^2(C_{s_1}^{l_1}+C_{s_2}^{l_2}-C_{s_3}^{l_3})=\frac{-\mathcal{P}_{s_1}^{l_1}\mathcal{P}_{s_2}^{l_2}+\mathcal{P}_{s_3}^{l_3}\mathcal{P}_{s_1}^{l_1}+4h^2+\mathcal{P}_{s_2}^{l_2}\mathcal{P}_{s_3}^{l_3}+4\Delta_{so}^2}
{\mathcal{P}_{s_1}^{l_1}\mathcal{P}_{s_2}^{l_2}\mathcal{P}_{s_3}^{l_3}+4h^2\mathcal{P}_{s_1}^{l_1}+4\Delta_{so}^2\mathcal{P}_{s_2}^{l_2}}. 
\label{eqn25}
\end{align}
Here, each $\mathcal{A}$ accounts for one set of coupled Cooperons, that is, one matrix equation from Eq.~\eqref{eqn24}}.

 The above equation is the main result of our work. \textcolor{black}{It is} readily evaluated analytically in the absence of the in-plane Zeeman field. 
 The divergent integral over momenta in Eq.~\eqref{eqn25} can be handled by introducing an upper cutoff associated with the inverse mean free path $l^{-1}=\sqrt{D\tau_0}$, which is the smallest length scale in our system. At $h=0$, we \textcolor{black}{then} obtain
	\begin{align}
	\frac{\delta\sigma}{\sigma_0}=&\,-2\Xi \ln\bigg(\frac{\tau^{-1}}{\tau_{\phi}^{-1}+\Gamma_x^x}\bigg)-\frac{1}{2}\ln \bigg(\frac{\tau^{-1}}{\tau_{\phi}^{-1}+\Gamma_z^0}\bigg)+\frac{1}{2}\ln \bigg(\frac{\tau^{-1}}{\tau_{\phi}^{-1}}\bigg)-
	\frac{1}{2}\ln \bigg(\frac{\tau^{-1}}{\tau_{\phi}^{-1}+\Gamma_0^z}\bigg)+
	\frac{1}{2}\ln \bigg(\frac{\tau^{-1}}{\tau_{\phi}^{-1}+\Gamma_z^z}\bigg)
	\nonumber \\ &+\gamma_{iv}\sum_{\pm}\pm \ln \bigg(\frac{\tau^{-1}}{\tau_{\phi}^{-1}+\Gamma_{iv}^{+}\pm\frac{\Gamma_{iv}^-}{\gamma_{iv}}}\bigg)
		+\Xi \gamma_{s} \sum_{\pm}\pm\ln \bigg(\frac{\tau^{-1}}{\tau_{\phi}^{-1}+\Gamma_{s}^{+}\pm\frac{\Gamma_{s}^-}{\gamma_{s}}}\bigg)
	. 
	\label{eqn26}
	\end{align}
\end{widetext}
	Here, we have introduced $\Gamma_{iv}^{\pm}=(\Gamma_x^z\pm \Gamma_x^0)\textcolor{black}{/2}$ and $\Gamma_s^{\pm}=(\Gamma_0^x\pm\Gamma_z^x)\textcolor{black}{/2}$, as well as
	\begin{equation}
	\gamma_{iv,s}=\frac{1}{\sqrt{1-\big(\frac{2\Delta_{so}}{\Gamma^-_{iv,s}}\big)^2}}.	
	\end{equation}
The coefficients $\gamma_{iv}$ and $\gamma_s$ capture the effect of the spin splitting. They are real if $1\geq 4\Delta_{so}^2/\Gamma_{iv,s}^2$, and imaginary otherwise. Although the rates $\Gamma^-_{iv,s}$ can be negative and the coefficients $\gamma_{iv,s}$ can be imaginary, their combination entering Eq.~\eqref{eqn26} is such that the imaginary parts cancel out, so that the conductance is always real (as it should be).  

Quantum interference is very sensitive to a magnetic field $B_\bot$ perpendicular to the monolayer, as it breaks the coherence of time\textcolor{black}{-}reversed paths of electrons, responsible for WL and WAL. This is used as a probe of W(A)L in experiments, which measure the magnetoconductance as a function of $B_\bot$. The perpendicular field couples to the momentum of the electrons, unlike the parallel field $B_\parallel$, which only couples to spin via the Zeeman effect. It leads to a quantization of momenta\textcolor{black}{,} $|\mathbf{Q}|\rightarrow Q_n=(n+1/2)/l_{B}^2$, where $n=0,1,2...$ denotes the Landau levels and $l_B=\sqrt{\hbar/4eB_\bot}$ is the magnetic length. We assume $l_B\gg l$, such that the diffusive limit is not violated, which imposes a constraint on the maximum field $B_\bot\ll \hbar/(4e D\tau_0)$. We \textcolor{black}{then} evaluate the magnetoconductance  $\Delta\sigma=\delta\sigma(B_\bot)-\delta\sigma(0)$ as
\begin{widetext}
	\begin{align}
\frac{\Delta\sigma}{\sigma_0}=&\, 2\Xi F\bigg(\frac{B_\bot}{B_{\phi}+B_x^x}\bigg)+\frac{1}{2}F \bigg(\frac{B_\bot}{B_{\phi}+B_z^0}\bigg)-\frac{1}{2} F \bigg(\frac{B_\bot}{B_{\phi}}\bigg)+
\frac{1}{2}F \bigg(\frac{B_\bot}{B_{\phi}+B_0^z}\bigg)-
\frac{1}{2} F\bigg(\frac{B_\bot}{B_{\phi}+B_z^z}\bigg)
\nonumber \\ &-\gamma_{iv} \sum_{\pm} \pm F \bigg(\frac{B_\bot}{B_{\phi}+B_{iv}^{+}\pm\frac{B_{iv}^-}{\gamma_{iv}}}\bigg)
-\Xi \gamma_{s} \sum_{\pm} \pm F \bigg(\frac{B_\bot}{B_{\phi}+B_{s}^{+}\pm\frac{B_{s}^-}{\gamma_{s}}}\bigg).
	\label{eqn28}
	\end{align}
\end{widetext}	
Here, we have introduced
\begin{equation}
F(z)=\ln(z)+\psi\bigg(\frac{1}{2}+\frac{1}{z}\bigg)\approx 
\begin{cases}
\frac{z^2}{24},\, &z\ll 1, \\
\ln z,\,&z\gg 1,
\end{cases}
\label{eqn29}
\end{equation}
where $\psi(z)$ is the digamma function, and  $B_i^j=\hbar\Gamma_i^j/(4eD)$ are effective magnetic fields associated with the scattering rates.

Eq.~\eqref{eqn28} acquires a simple form if the decoherence rate $\tau_\phi^{-1}$ is either the dominant or the smallest scattering rate. For very long $\tau_\phi$, such that  $\tau_{\phi}^{-1}\ll \Gamma_s^l$, all the gapped Cooperons can be neglected, and only the third term in Eq.~\eqref{eqn28} remains. Then, we have   $\Delta\sigma/\sigma_0=-\textcolor{black}{(} 1/2 \textcolor{black}{)}F(B_\bot/B_\phi)$, as in conventional metal with strong \textcolor{black}{spin-dependent} disorder. For short decoherence times, $\tau_{\phi}^{-1}\gg \Gamma_s^l$, all the Cooperon gaps can be neglected. Different contributions to Eq.~\eqref{eqn28} then cancel pairwise, and we obtain $\Delta\sigma/\sigma_0=2\Xi F(B_\bot/B_\phi)$. This exhibits WL, WAL or a vanishing quantum correction for $\Xi=1, -1,0 $ respectively, similarly to a Dirac material in a smooth disorder potential. \textcolor{black}{This limiting case contributes to the interference correction with a four times larger prefactor compared to the previous one - a consequence of spin and valley degeneracy. }

The magnetoconductance formula Eq.~\eqref{eqn28} captures the rich \textcolor{black}{weak} localization behavior of TMDCs and graphene/TMDC. Due to the large number of parameters it is difficult to apply it directly to experiments. In the next section, we will present and discuss several realistic regimes in which this result significantly simplifies, and compare them to the existing theories. \textcolor{black}{Furthermore, we will discuss the effect of a finite in-plane Zeeman field.}

\section{Discussion}\label{sec4}
We will proceed by analyzing the magneto\textcolor{black}{conductance} formula \eqref{eqn28} in the regimes of strong (Sec.~\ref{subsecA}) and weak short-range disorder (Sec.~\ref{subsecB}). We will also address the effect of an in-plane Zeeman field (Sec.~\ref{subsecC}).

\subsection{Strong short-range disorder}\label{subsecA}
The regime where intervalley scattering dominates over all spin-dependent scattering rates,  $\tau_{iv}^{-1}\gg \tau_{i,j}^{-1}$, with $i=z,zv,iv$ and $j=z,o$, is the most commonly used regime when interpreting the measurements of the quantum correction. \textcolor{black}{Such a large magnitude of intervalley scattering is expected in samples with an abundance of atomic defects, or in small samples, where the edges can contribute to this kind of scattering.} \textcolor{black}{In that case,} the effect of spin-dependent disorder can be captured with only two scattering rates,
\begin{align}
&\textcolor{black}{\tau_{sym}^{-1}=2(\tau_{z,e}^{-1}+\tau_{zv,e}^{-1}+\tau_{iv,e}^{-1})}, \nonumber \\ 
&\tau_{asy}^{-1}=\tau_{z,o}^{-1}+\tau_{zv,o}^{-1}+\tau_{iv,o}^{-1}+\tau_{BR}^{-1} 
\label{eqn30}.
\end{align}
Here $\tau_{sym}^{-1}$ contains all the spin-dependent scattering processes that satisfy mirror ($z\rightarrow -z$) symmetry and\textcolor{black}{,} thus\textcolor{black}{,} preserve the electron spin. On the other hand, $\tau_{asy}^{-1}$ contains spin-flip processes that break this symmetry. \textcolor{black}{In the presence of potential disorder only, we can use the estimates provided in Table \ref{table1} to identify \textcolor{black}{the} dominant contributions to these rates. In that case, we find that the symmetric rate is dominated by $\tau_{z,e}^{-1}$, which describes the Elliott-Yafet spin-relaxation mechanism induced by Kane-Mele SOC, while the asym\textcolor{black}{m}etric rate is dominated by $\tau_{BR}^{-1}$, which describes the Dyakonov-Perel spin relaxation mechanism induced by Rashba SOC. If additional \textcolor{black}{spin-orbit} impurities are present in the system, the symmetric and asymmetric rates are not limited by the band structure SOC parameters.}

 In this regime, $\Gamma_{iv}^-\approx\Gamma_{iv}^+\approx \tau_{iv}^{-1}$, and $\gamma_{iv}\approx 1/\sqrt{1-4\Delta_{so}^2\tau_{iv}^2}$. Furthermore, we will assume that the effect of trigonal warping captured in  $\tau_*^{-1}$ and $\tau_{**}^{-1}$ for intravalley Cooperons (see the bottom of Table \ref{table2}) is small compared to intervalley scattering. \textcolor{black}{Then,} we have $\tau_{*}^{-1}\approx\tau_{**}^{-1}\approx\tau_{iv}^{-1}$, and the magnetoconductance \eqref{eqn28} becomes
\begin{widetext}
\begin{align}
\frac{\Delta\sigma}{\sigma_0}=& \,2\Xi F\bigg(\frac{B_\bot}{B_\phi+B_{iv}}\bigg)+\frac{1}{2}F\bigg(\frac{B_\bot}{B_\phi+2 B_{asy}}\bigg)-\frac{1}{2}F\bigg(\frac{B_\bot}{B_\phi}\bigg) \nonumber \\
&- \gamma_{iv}\bigg[  \textcolor{black}{F\bigg(\frac{B_\bot}{B_\phi+B_{iv}(1+\frac{1}{\gamma_{iv}})}\bigg)}-
F\bigg(\frac{B_\bot}{B_\phi+B_{iv}(1-\frac{1}{\gamma_{iv}})+B_s}\bigg)
\bigg].
\label{eqn31}
\end{align}
\end{widetext}
 Here $\tau_s^{-1}=\tau_{sym}^{-1}+\tau_{asy}^{-1}$, and $B_i=\hbar/(4eD\tau_i)$. We see that the magnetoconductance is determined by \textcolor{black}{a} combination of valley and spin physics,  described by the intervalley scattering rate $\tau_{iv}^{-1}$, and spin scattering rates $\tau_{sym}^{-1}$ and $\tau_{asy}^{-1}$.  The interplay between intervalley scattering and valley-dependent SOC is captured by the coefficient $\gamma_{iv}$. We will proceed by analyzing this interplay in two limits: $\tau_{iv}^{-1}\gg \Delta_{so}$ and $\Delta_{so}\gg \tau_{iv}^{-1}$.
 
 Within these two limits, we can readily address 3 regimes of the decoherence rate: $(i)\, \tau_\phi^{-1}\ll \tau_s^{-1}$, $(ii) \,\tau_s^{-1}\ll \tau_\phi^{-1}\ll \tau_{iv}^{-1}$, and $ (iii)\, \tau_{iv}^{-1}\ll \tau_\phi^{-1}$, where the quantum correction acquires a simple form. The cases $(i)$ and $(iii)$, where the decoherence rate is the dominant or the smallest one, respectively, were previously discussed in the general context of Eq.~\eqref{eqn28}. The intermediate regime $(ii)$ is not universal. In the limit $\tau_{iv}^{-1}\gg \Delta_{so}$, it yields $\Delta\sigma/\sigma_0=F(B/B_\phi)$. This is analogous to \textcolor{black} {a} conventional metal without SO impurities, and represents a sum of three spin-triplets $C_i^0$ ($i=x,y,z$), which contribute as $\textcolor{black}{(}3/2\textcolor{black}{)} F(B_\bot/B_\phi)$, and a spin-singlet $C_0^0$, which contributes as $-\textcolor{black}{(}1/2\textcolor{black}{)}F(B_\bot/B_\phi)$. For $\Delta_{so}\gg \tau_{iv}^{-1}$, the two triplets $C_x^0$ and $C_y^0$ are suppressed by the SOC, and the quantum correction vanishes. 
 
  We obtain more complex behavior in the crossover regimes  $\tau_\phi^{-1}\sim \tau_{s}^{-1}$ [which includes $(i)$ and $(ii)$] and $\tau_\phi^{-1}\sim\tau_{iv}^{-1}$ [which includes $(ii)$ and $(iii)$]. Strong intervalley scattering completely suppresses the valley structure in the first regime, so that the magnetoconductance is determined by the spin physics only. On the other hand, the valley physics dominates in the second regime, as the effect of spin-scattering is washed out by electron decoherence.

\paragraph*{\textcolor{black}{a. \, }Limit $\tau_{iv}^{-1}\gg \Delta_{so}$: \label{a}} Here, Eq.~\eqref{eqn31} simplifies, as $\gamma_{iv}\approx 1$. 
In the crossover regime $\tau_\phi^{-1}\sim \tau_{s}^{-1}$, the first and the \textcolor{black}{fourth} term of Eq.~\eqref{eqn31} are suppressed by the large intervalley scattering, and we obtain
 \begin{equation}
 \frac{\Delta\sigma}{\sigma_0}=\frac{1}{2}F\bigg(\frac{B_\bot}{B_\phi+2B_{asy}}\bigg)-\frac{1}{2}F\bigg(\frac{B_\bot}{B_\phi}\bigg)+ F\bigg(\frac{B_\bot}{B_\phi+\tilde{B}_s}\bigg),
 \label{eqn32}
 \end{equation}
 Here, we have introduced 
 \begin{equation}
 \textcolor{black}{\tilde{\tau}_s^{-1}= \tau_{iv}^{-1}\bigg(1-\frac{1}{\gamma_{iv}}\bigg)+\tau_s^{-1} \approx 2\Delta_{so}^2\tau_{iv}+\tau_s^{-1},}
 \end{equation}
and $\tilde{B}_s=\hbar/(4 e D \tilde{\tau}_s)$. \textcolor{black}{As valley structure and spin-splitting are suppressed in this regime, the system behaves similarly to a diffusive metal with spin-orbit impurities, and Eq.~\eqref{eqn32} is equivalent to the HLN formula. This remains true even when intervalley scattering becomes comparable to intervalley scattering, for $\tau_{iv}^{-1}\sim \tau_0^{-1}$. The equation \eqref{eqn32} still holds in that case, although with a modified diffusion constant (see Appendix~\ref{ap2}).}
 
  The effect of valley-dependent SOC is captured by an additional contribution to the symmetric rate, $\tau_{sym}^{-1}\rightarrow \tau_{sym}^{-1}+2\Delta_{so}^2\tau_{iv}$,
\textcolor{black}{which stems from the coupling of the Cooperon modes $C_x^{0(z)}$ with $C_y^{z(0)}$  by this SOC}.
 This effect was already discussed in Refs.~\onlinecite{cummings2017giant,zihlmann2017large,garcia2018spin}\textcolor{black}{,} and used to estimate $\Delta_{so}$ from the experimental data in graphene/TMDC heterostructures.  However, the estimated SOC is of the same order of magnitude as $\tau_{iv}^{-1}$, which is outside of the region of validity of this formula ($\tau_{iv}^{-1}\gg \Delta_{so}$). Instead, the full formula provided by Eq.~\eqref{eqn31} should be used in order to get a more reliable estimate of the valley-dependent SOC. 
 
 \textcolor{black}{If $\tilde{\tau}_s^{-1}\sim \tau_\phi^{-1}\sim \tau_{asy}^{-1}$, Eq.~\eqref{eqn32} exhibits WAL-WL crossover as the magnitude of the perpendicular field is increased.}
 \textcolor{black}{We next consider the regime $\tilde{\tau}_s\gg \tau_\phi^{-1}\sim \tau_{asy}^{-1}$. Here, the last term of Eq.~\eqref{eqn32} is suppressed due to the combined effect of all mirror-symmetric SOC in the system, as $\tau_{sym}^{-1}+2\Delta_{so}^2\tau_{iv}\gg\tau_\phi^{-1}.$ We thus have
  \begin{equation}
 \frac{\Delta\sigma}{\sigma_0}=\frac{1}{2}F\bigg(\frac{B_\bot}{B_\phi+2 B_{asy}}\bigg)-\frac{1}{2}F\bigg(\frac{B_\bot}{B_\phi}\bigg) .
 \label{eqn34}
 \end{equation}
This corresponds to pure WAL behavior as a function of $B_\bot$, that saturates on the scale of $B_{asy}$. This kind of saturation was noticed in several recent experiments that show flat WAL curves, such as Refs. \onlinecite{wakamura2017strong, costanzo2016gate, zihlmann2017large}}.
 \textcolor{black}{The interference correction vanishes for  $\tilde{\tau}_s^{-1}\gg\tau_\phi^{-1}\gg \tau_{asy}^{-1}$, and shows pure WL behavior if $\tilde{\tau}_s^{-1}\sim \tau_\phi^{-1}\gg \tau_{asy}^{-1}$, given as}
 \textcolor{black}{
 \begin{equation}
\frac{\Delta\sigma}{\sigma_0}=F\bigg(
\frac{B_{\bot}}{B_\phi+\tilde{B}_s}
\bigg).
 \end{equation}
}

Next, we address the crossover regime $\tau_{\phi}^{-1}\sim \tau_{iv}^{-1}$. \textcolor{black}{Here}, the spin scattering rates can be neglected, and the second and third term of Eq.~\eqref{eqn31} cancel out, which yields
 \begin{equation}
 \frac{\Delta\sigma}{\sigma_0}=2\Xi F\bigg(\frac{B_\bot}{B_\phi+B_{iv}}\bigg)+F\bigg(\frac{B_\bot}{B_\phi}\bigg)-F\bigg(\frac{B_\bot}{B_\phi+2B_{iv}}\bigg) 
 \label{eqn33}.
 \end{equation}
\textcolor{black}{This result at $\Xi=-1$ is equivalent to Ref.~\onlinecite{mccann2006weak}, which describes graphene without spin-dependent impurities}. As a function of a perpendicular field, it exhibits pure WL for $\Xi=1$ and $\Xi=0$, and \textcolor{black}{a} WL-WAL crossover for $\Xi=-1$.

\begin{figure}[h!]
	\includegraphics[width=0.45\textwidth]{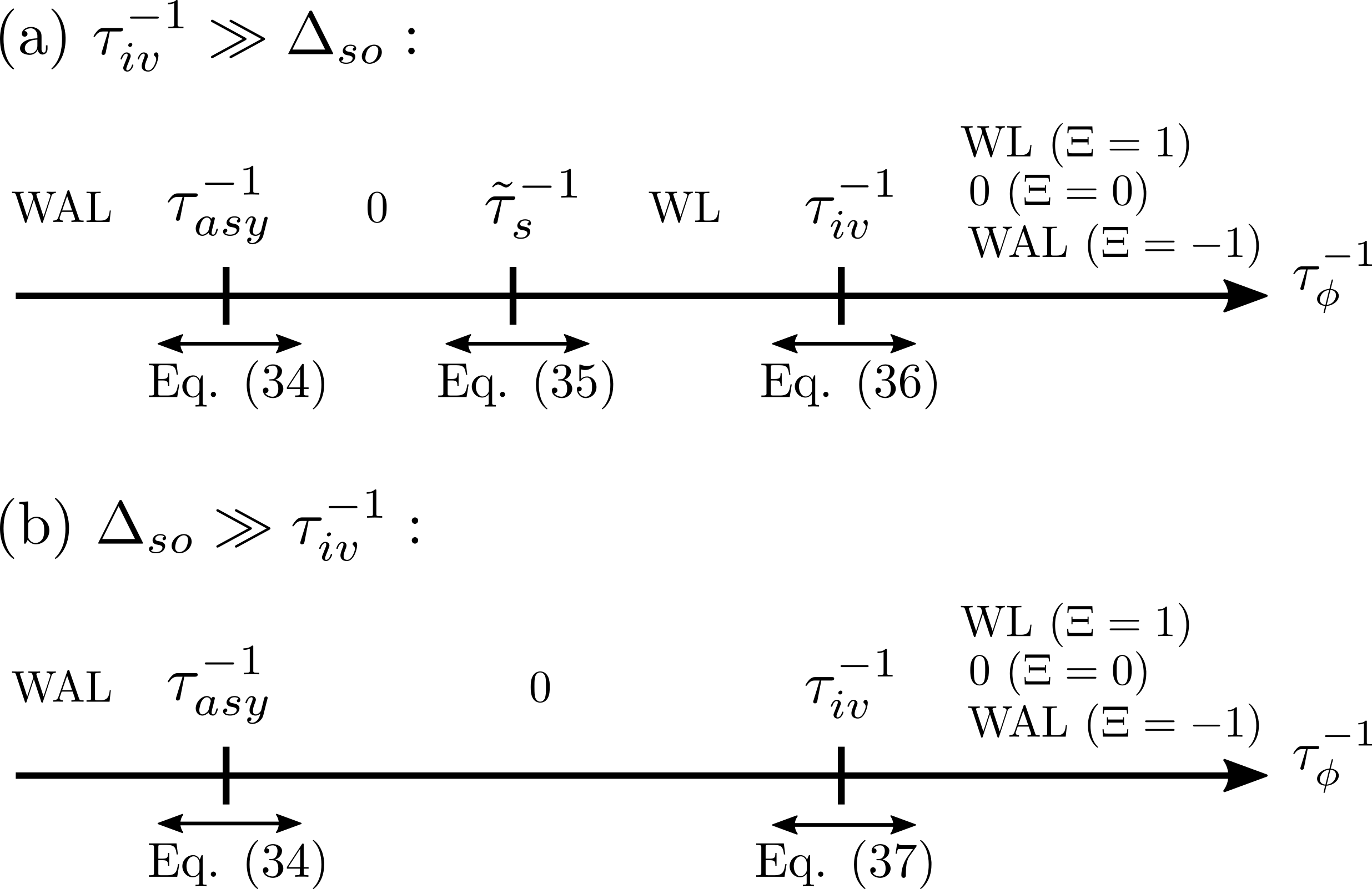}
	\caption{Schematic representation of the  WL behavior in the regime of strong short-range disorder, $\tau_{iv}^{-1}\gg \tau_s^{-1}$. In the crossover regions described by Eqs.~\eqref{eqn32}-\eqref{eqn35}, the magnetoconductance at low (high) perpendicular field behaves the same as in the left (right) adjacent region on the $\tau_\phi^{-1}$ arrow. \textcolor{black}{In panel (a), the regime of vanishing interference correction between $\tau_{asy}^{-1}$ and $\tilde{\tau}_s^{-1}$ disappears if $\tau_{asy}^{-1} \sim \tilde{\tau}_{sym}^{-1}$}. }
	\label{figure3}
\end{figure}

 Fig.~\ref{figure3}(a) gives a schematic representation of the different regimes in the limit $\tau_{iv}^{-1}\gg \Delta_{so}$.

\begin{figure*}[t!]
	\includegraphics[width=\textwidth]{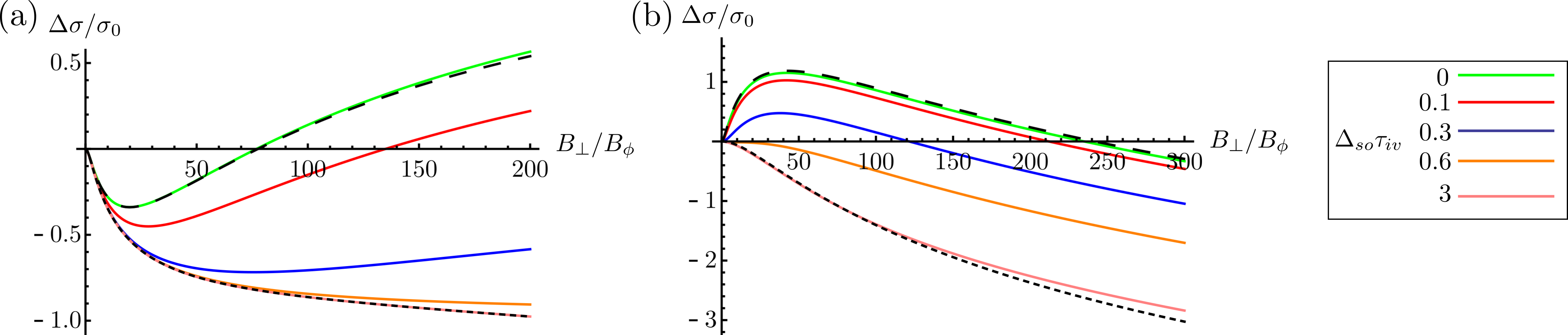}
	\caption{Interference-induced magnetoconductance as a function of a weak perpendicular magnetic field under the influence of increasing valley-dependent SOC. We take the chemical potential to be deep in the conduction band, such that $\Xi=-1$. \textcolor{black}{The fields $B_{sym}$ and $B_{asy}$ are determined by the Elliott-Yaffet contribution from the Kane-Mele SOC, and the Dyakonov-Perel contribution due to the Rashba SOC, respectively, as well as other sources of spin-orbit scattering [see Table~\ref{table1} and Eq.~\eqref{eqn30}]. The effect of the valley-Zeeman SOC is captured by the parameter $\Delta_{so}\tau_{iv}$}. (a) All curves are plotted for the parameters $B_{iv}=200 B_{\phi}, B_{sym}=B_{asy}=3 B_{\phi}$. The dashed black line corresponds to Eq.~\eqref{eqn32}, while the dotted line corresponds to Eq.~\eqref{eqn34} (b) All curves are plotted for the parameters $B_{iv}=10B_\phi, B_{sym}=B_{asy}=0.02 B_{\phi}$. The dashed black line corresponds to Eq.~\eqref{eqn33}, while the dotted line corresponds to Eq.~\eqref{eqn35}.}
	\label{figure4}
\end{figure*}

 \paragraph*{\textcolor{black}{b. \, }Limit $\Delta_{so}\gg \tau_{iv}^{-1}$: \label{b}}
 Since $\gamma_{iv}\approx 0$, \textcolor{black}{here} only the first three terms of Eq.~\eqref{eqn31} contribute to the magnetoconductance. In the crossover regime  $\tau_{\phi}^{-1}\sim \tau_s^{-1}$,\textcolor{black}{we again obtain Eq.~\eqref{eqn34}. Similarly to the previously considered case analyzed below Eq.~\eqref{eqn34}, saturated WAL in this regime can be understood as a consequence of strong mirror-symmetric SOC which suppresses Cooperons that would lead to WL. However, this suppression is now predominantly caused by spin-splitting due to $\Delta_{so}$, irrespective of the magnitude of $\tau_{sym}^{-1}$. This regime, therefore, presents an alternative to the standard HLN theory to interpret the experiments showing saturated WAL signals.}

 Finally, we analyze the crossover regime  $\tau_{\phi}^{-1}\sim \tau_{iv}^{-1}$. We find
 \begin{equation}
 \frac{\Delta\sigma}{\sigma_0}=2\Xi F\bigg(\frac{B_\bot}{B_\phi+B_{iv}}\bigg),
 \label{eqn35}
 \end{equation}
 which exhibits pure WAL, pure WL, or vanishes for $\Xi=1$, $\Xi=-1$ and $\Xi=0$, respectively.
 
  Fig.~\ref{figure3}(b) gives a schematic representation of the different regimes in the limit $\Delta_{so}\gg \tau_{iv}^{-1}$.

 Fig.~\ref{figure4} illustrates the behavior of the magnetoconductance beyond the two extreme limits $\tau_{iv}^{-1}\gg \Delta_{so}$ and $\Delta_{so}\gg \tau_{iv}^{-1}$, analyzed above. \textcolor{black}{In particular}, Fig.~\ref{figure4}(a) addresses the crossover from the regime described by Eq.~\eqref{eqn32} to Eq.~\eqref{eqn34} as the magnitude of valley-dependent SOC is increased. Similary  Fig.~\ref{figure4}(b) shows a crossover from Eq.~\eqref{eqn33} to Eq.~\eqref{eqn35}.

 \subsection{Weak short-range disorder}\label{subsecB}
 In this section\textcolor{black}{,} we analyze the regime where intervalley scattering rate is much weaker than the \textcolor{black}{spin-scattering rates, $\tau_{sym}^{-1},\tau_{asy}^{-1}\gg \tau_{iv}^{-1}$}, \textcolor{black}{which is appropriate for large samples without atomic defects}. The intervalley spin-scattering rates are assumed to be even weaker, $ \tau_{iv,e/o}^{-1} \ll \tau_{iv}^{-1}$, and thus neglected. The magnetoconductance formula is then given as
 \begin{align}
&\frac{\Delta\sigma}{\sigma_0}=\,2\Xi F\bigg(\frac{B_\bot}{B_{\phi}+B_x^x}\bigg)-\frac{1}{2} F\bigg(\frac{B_\bot}{B_{\phi}}\bigg) 
\nonumber\\&+
\frac{1}{2} F\bigg(\frac{B_\bot}{B_{\phi}+2 B_{iv}}\bigg)
-\Xi \gamma_{s}\sum_{\pm}\pm F \bigg(\frac{B_\bot}{B_{\phi}+B_{s}^{+}\pm\frac{B_{s}^-}{\gamma_{s}}}\bigg)
.
\label{eqn36}
 \end{align}
In this regime, the quantum correction  is governed by the interplay between $\Delta_{so}$ and a combination of the spin-scattering rates $\Gamma_s^{-}$, described by the coefficient $\gamma_s$. Unlike the case of strong short-range disorder, the Cooperons containing $\gamma_{iv}$ cancel out in this regime, so the ratio of intervalley scattering and valley-dependent SOC does not affect $\Delta\sigma$. The three intravalley Cooperon gaps  \textcolor{black}{$\Gamma_i^x$ $(i=0,x,y,z)$} that enter Eq.~\eqref{eqn36} \textcolor{black}{have a similar structure}. To simplify further analysis, we will assume that they are of the same order of magnitude.

We proceed similarly to the previous section, and analyze the \textcolor{black}{three} extreme limits with respect to the decoherence rate. If it is the smallest, $\tau_\phi^{-1}\ll \tau_{iv}^{-1}$, or the largest, $\textcolor{black}{\Gamma_i^x}\ll\tau_\phi^{-1}$, scattering rate, the general arguments presented after Eq.~\eqref{eqn28} apply. In the intermediate limit $\tau_{iv}^{-1}\ll\tau_\phi^{-1}\ll\textcolor{black}{\Gamma_i^x}$, the quantum correction vanishes.

\begin{figure}[h!]
  	\includegraphics[width=0.45\textwidth]{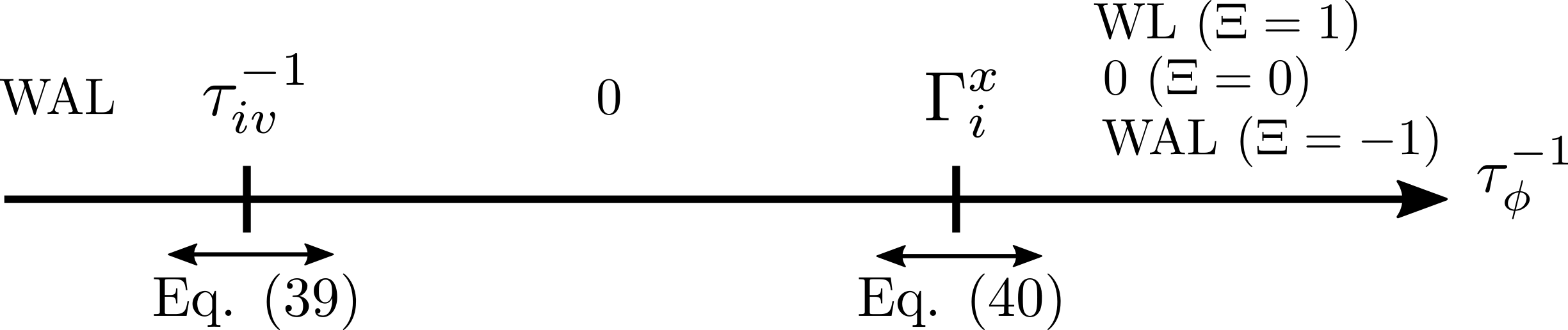}
  	\caption{Schematic representation of the WL behavior in the regime of weak short-range disorder, \textcolor{black}{$\tau_{sym}^{-1},\tau_{asy}^{-1}\gg \tau_{iv}^{-1}\gg \tau_{iv,e/o}^{-1}$}. The behavior in the crossover regions is represented in the same way as in  Fig.~\ref{figure3}.} 
  	\label{figure5}
  \end{figure}

We next examine the crossover regimes. For $\tau_{\phi}^{-1} \sim \tau_{iv}^{-1}$, we have
\begin{equation}
\frac{\Delta\sigma}{\sigma_0}=-\frac{1}{2}F\bigg(\frac{B_\bot}{B_\phi}\bigg)+\frac{1}{2}F\bigg(\frac{B_\bot}{B_\phi+2B_{iv}}\bigg).
\label{eqn37}
\end{equation}
This formula is determined by intervalley scattering only, and exhibits WAL behavior which saturates on the scale of $B_{iv}$.
Finally, in the crossover regime $\tau_{\phi}^{-1}\sim \textcolor{black}{\Gamma_i^x}$ we have
\begin{align}
\frac{\Delta\sigma}{\sigma_0}=&\, 2\Xi F\bigg(\frac{B_\bot}{B_{\phi}+B_x^x}\bigg) \nonumber \\
&- \Xi\gamma_s\sum_{\pm}\pm F \bigg(\frac{B_\bot}{B_{\phi}+B_s^+\pm\frac{B_s^-}{\gamma_s}}\bigg)
.
\label{eqn38}
\end{align}
 In the limit $\Gamma_s^-\gg \Delta_{so}$, one should consider all three terms in Eq.~\eqref{eqn38} since $\gamma_s\approx 1$. As $\Delta_{so}$ increases, the second line of Eq.~\eqref{eqn38} becomes suppressed, until it vanishes for  $\Delta_{so}\gg\Gamma_s^-$, where $\gamma_s\approx 0$. We see that the qualitative behavior of the magnetoconductance remains the same for any $\gamma_s$, and thus, any $\Delta_{so}$. It only depends on the doping coefficient $\Xi$, and exhibits WL, WAL\textcolor{black}{,} or neither for $\Xi=1, -1$\textcolor{black}{,} and $0$, respectively. These conclusions are schematically represented in Fig.~\ref{figure5}.

\begin{figure*}[]
	\includegraphics[width=\textwidth]{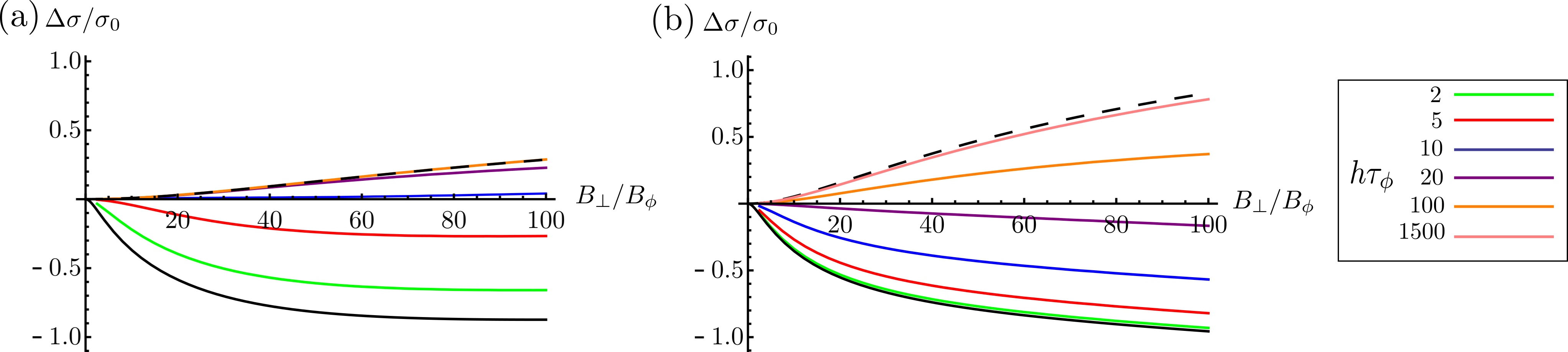}
	\caption{Influence of the in-plane Zeeman field on the magnetoconductance curves. The solid black line represents the curve at zero in-plane Zeeman field, while the dashed line represents the saturation curve given by Eq.~\eqref{eqn40} at high fields. (a)  The parameters for the plot are $B_{iv}=100 B_{\phi}$, $B_{sym}=B_{asy}=10 B_{\phi}, B_{so}=0$\textcolor{black}{,} and $\Xi=-1$. The crossover to WL happens at $B_\bot\approx 10 B_{\phi}$. (b) The parameters for the plot are $B_{iv}=100 B_{\phi}$, $B_{sym}=B_{asy}=3.5 B_{\phi}, B_{so}=120 B_{\phi}$\textcolor{black}{,} and $\Xi=-1$. The crossover to WL happens at $B_\bot\approx 30 B_\phi.$}
	\label{figure6}
\end{figure*} 
\subsection{Influence of the in-plane Zeeman field}\label{subsecC}

One of the main difficulties when experimentally extracting the parameters from quantum magneto\textcolor{black}{conductance} fits comes from the fact that there are multiple parameter combinations that can fit the same data. For example, both valley-dependent SOC and spin-dependent scattering can lead to pronounced WAL signals. Applying an in-plane Zeeman field can help overcome these ambiguities, as different kinds of disorder and SOC interplay differently with the field. 

At sufficiently high in-plane Zeeman field, all spin-singlet $C_0^{\textcolor{black}{l}}$ and spin-triplet $C_x^{\textcolor{black}{l}}$ \textcolor{black}{Cooperons} are suppressed, and we arrive at the asymptotic formula for the magnetoconductace\textcolor{black}{,}
\begin{align}
\frac{\Delta\sigma}{\sigma_0}=&\sum_{i=x,z} \bigg[
\Xi F\bigg(\frac{B_\bot}{B_{\phi}+B_i^x}\bigg) \nonumber  \\
&+\frac{1}{2}F\bigg(\frac{B_\bot}{B_{\phi}+B_i^0}\bigg)-
\frac{1}{2}F\bigg(\frac{B_\bot}{B_{\phi}+B_i^z}\bigg)
\bigg].
\label{eqn39}
\end{align}
The magnitude of the in-plane Zeeman field required to reach the high-field formula \eqref{eqn39} differs depending on the parameter regime, as will be discussed in the following. Note that it will always be reached if  $h \gg \Delta_{so}, \tau_i^{-1}$, where $\tau_i^{-1}$ are all scattering rates except the diagonal one\textcolor{black}{,} $\tau_0^{-1}$. 

First, we analyze the regime \textcolor{black}{where the} short-range disorder \textcolor{black}{rate is much larger than all spin-dependent disorder rates},  $\tau_{iv}^{-1}\gg \tau_s^{-1}$. In this case the asymptotic formula  acquires the form
\begin{align}
\frac{\Delta\sigma}{\sigma_0}=& \,2\Xi F\bigg(\frac{B_\bot}{B_\phi+B_{iv}}\bigg)+\frac{1}{2}F\bigg(\frac{B_\bot}{B_\phi+2B_{asy}}\bigg) \nonumber \\
&+\frac{1}{2}F\bigg(\frac{B_\bot}{B_\phi+B_{s}}\bigg)- F\bigg(\frac{B_\bot}{B_\phi+2 B_{iv}}\bigg)
\label{eqn40}.
\end{align}
Starting from the general expression~\eqref{eqn25}, we will next check the magnitude of $h$ needed to reach this formula in the limits $\tau_{iv}^{-1}\gg \Delta_{so}$ and $\Delta_{so}\gg \tau_{iv}^{-1}$. 

Let us consider $\tau_{iv}^{-1}\gg \Delta_{so}$. \textcolor{black}{If the decoherence rate $\tau_\phi^{-1}$ is larger than all spin-scattering rates, the spin structure is suppressed, and the in-plane Zeeman field has no effect. In this case, the formula \eqref{eqn40} is valid for any $h$ and is equivalent to Eq.~\eqref{eqn33}. On the other hand, if $\tau_\phi^{-1}$ is of the order of the spin-scattering rates, all the valley-singlet Cooperons, $C^0_s$,  contribute to the magnetoconductance at $h=0$ [Eq.~\eqref{eqn32}], and a finite $h$ acts by suppressing  the spin-singlet Cooperon $C_0^0$ and the spin-triplet  Cooperon $C_x^0$. For fields of the order $\tilde{\tau}_s^{-1}\ll h\ll \tau_{iv}^{-1}$, Eq.~\eqref{eqn40} holds, but with $B_s$ replaced with $\tilde{B}_s$. Therefore, unless $\tau_s^{-1}\gg \Delta_{so}^2\tau_{iv}$, the valley-dependent SOC still has an effect at such fields, through the contribution $2\Delta_{so}^2\tau_{iv}$ to the effective rate $\tilde{\tau}_s^{-1}$. In that case, the high-field asymptotic formula is reached only at very high fields of the order of intervalley scattering, namely $h\gg \tau_{iv}^{-1}$}.

Next, we consider the limit  $\Delta_{so}\gg \tau_{iv}^{-1}$. In this regime, the Cooperons $C_i^j$ and $C_j^i$, where $i=x,y$ and $j=0,z$, are suppressed by the strong $\Delta_{so}$ at $h=0$. In order to reach the asymptotic formula Eq.~\eqref{eqn40}, a large field $h\gg \Delta_{so}$ is needed. It negates the effect of the valley-dependent SOC and restores $C_y^j$ and $C_z^{i}$ \textcolor{black}{Cooperons}, while  suppressing all $C_0^{\textcolor{black}{l}}$ and $C_x^{\textcolor{black}{l}}$ \textcolor{black}{Cooperons}.

Finally, we address the limit of weak short-range disorder, \textcolor{black}{$\tau_{sym}^{-1},\tau_{asy}^{-1}\gg \tau_{iv}^{-1}\gg \tau_{iv,e/o}^{-1}$}, described by Eq.~\eqref{eqn36} at $h=0$.  Similarly to the previously considered case, strong $h$ negates the effect of $\Delta_{so}$ and suppresses all spin-singlet and $x$-triplet \textcolor{black}{Cooperons}. Here, the asymptotic formula takes the form
\begin{equation}
\frac{\Delta\sigma}{\sigma_0}=\Xi \sum_{i=x,z} F\bigg(\frac{B_\bot}{B_{\phi}+B_{i}^x}\bigg),
\label{eqn41}
\end{equation}
and is reached if the in-plane Zeeman field is the largest energy scale, $h\gg \textcolor{black}{\Gamma_i^x},\Delta_{so}, \tau_{\phi}^{-1}$ \textcolor{black}{($i=0,x,y,z$)}. The prefactor $\Xi$ indicates that it can exhibit WAL, WL\textcolor{black}{,} or neither depending on the doping, similarly to Eq.~\eqref{eqn38}. 

 To illustrate a situation where applying the in-plane field can help in the interpreta\textcolor{black}{t}ion of the quantum correction, we plot two magnetoconductance curves with a similar shape, but with significantly different parameters in
Fig.~\ref{figure6} (black line). The first curve [Fig.~\ref{figure6}(a)] has strong spin-scattering and no valley-dependent SOC, while the second one has weaker spin-scattering and strong SOC [Fig.~\ref{figure6}(b)]. The high-field saturation curve (dashed line) has a similar shape in both cases, and is described by Eq.~\eqref{eqn40}. The amplitude of WL at high fields is \textcolor{black}{somewhat} larger in the case of strong SOC, as the spin-orbit scattering is weaker, which means that the second line of Eq.~\eqref{eqn40} gives a larger contribution compared to the other case. More importantly, this case is more resistant to the effect of the applied field, and the crossover to WL  happens at a much higher field amplitude. This is consistent with the above analysis, as the expected crossover field is $h\sim \tau_s^{-1}$ for Fig.~\ref{figure6}(a) and  $h\sim \Delta_{so}$ for Fig.~\ref{figure6}(b). Thus, applying an in-plane field helps distinguish the contributions of valley-dependent SOC and spin-dependent scattering to the quantum correction.

 \section{Conclusions}
 In conclusion, we have developed a theory of weak localization and magnetoconductance for TMDC monolayers and their heterostructures with graphene, using \textcolor{black}{the} standard diagrammatic technique for disordered systems. The interplay between spin and valley physics in these materials yields a rich behavior of the quantum correction to the conductivity, which we discuss in several regimes of interest for the interpretation of recent experimental data. We generalize the HLN and MF theories and propose a formula that can be used to extract the magnitude of valley-dependent SOC and disorder from the experiments in all regimes. In some cases, interpreting the experiments is not straightforward, as different parameter combinations \textcolor{black}{may} explain the data equally well. An in-plane Zeeman field can be used as an additional tuning parameter to help distinguish between the contributions of different processes. 
 
 \begin{acknowledgments}
 
We thank T. Wakamura, H. Bouchiat, and S. Gueron for helpful discussions. We acknowledge funding from the Laboratoire d'excellence LANEF in Grenoble (ANR-10-LABX-51-01)\textcolor{black}{,} and by the ANR through grants No. ANR-16-CE30-0019 and ANR-17-PIRE-0001.
 \end{acknowledgments}
\appendix
\textcolor{black}{\section{Higher-order corrections due to the valley-dependent SOC and in-plane Zeeman field \label{ap1}}}
\textcolor{black}{
As discussed in Sec.~\ref{subsec2}, the main effect of the spin-splitting fields, $h$ and $\Delta_{so}$, is the coupling of different Cooperon modes. However, it is also important to consider the corrections beyond the leading order in $\tau_0$ in the diffusive limit, by keeping the terms of the order  $\Delta_{so}^2\tau_0$, $h^2\tau_0$, and $h\Delta_{so}\tau_0$, as they can be of comparable magnitudes to the scattering rates appearing in the Cooperon gaps. In this Appendix, we will discuss these corrections, and show that they can always be neglected when compared to the leading-order effect of $h$ and $\Delta_{so}$. }

\textcolor{black}{Firstly, we generalize Eq.~\eqref{eqn24} to include these corrections. We have
 \begin{widetext}
 	\begin{align}
 	\begin{bmatrix}
 	\mathcal{P}_0^{x(y)}+4\rho^2\tau_0 & \mp 2\Delta_{so} & -2ih \\
 	\pm 2\Delta_{so} & \mathcal{P}_z^{y(x)}+4\Delta_{so}^2\tau_0 & \pm 4ih\Delta_{so}\tau_0 \\
 	-2ih & \mp 4ih\Delta_{so}\tau_0 & \mathcal{P}_x^{x(y)}+4h^2\tau_0
 	\end{bmatrix}
 	\begin{bmatrix}
 	C_{00}^{xx(yy)} & C_{0z}^{xy(yx)} & C_{0x}^{xx(yy)}\\
 	C_{z0}^{yx(xy)} & C_{zz}^{yy(xx)} & C_{zx}^{yx(xy)} \\
 	C_{x0}^{xx(yy)} & C_{xz}^{xy(yx)} & C_{xx}^{xx(yy)}
 	\end{bmatrix}= \frac{1}{2\pi\nu\tau_0^2},\nonumber \\
 	\begin{bmatrix}
 	\mathcal{P}_x^{0(z)} +4\rho^2\tau_0& -2\Delta_{so} & -2 i h \\
 	2\Delta_{so} & \mathcal{P}_y^{z(0)}+4\Delta_{so}^2\tau_0 & 4 i h \Delta_{so}\tau_0 \\
 	-2 i h & -4 i h \Delta_{so}\tau_0 & \mathcal{P}_0^{0(z)}+4h^2\tau_0
 	\end{bmatrix}
 	\begin{bmatrix}
 	C_{xx}^{00(zz)} & C_{xy}^{0z(z0)} & C_{x0}^{00(zz)}\\
 	C_{yx}^{z0(0z)} & C_{yy}^{zz(00)} & C_{y0}^{z0(0z)}\\
 	C_{0x}^{00(zz)} & C_{0y}^{0z(z0)} & C_{00}^{00(zz)}
 	\end{bmatrix}=\frac{1}{2\pi\nu \tau_0^2},
 	\label{eqna1}
 	\end{align}
\end{widetext}
where $\rho^2=\Delta_{so}^2+h^2$. We see that all the gaps related to the Cooperons coupled by the in-plane Zeeman field, $\Gamma_0^l$ and $\Gamma_x^l$, are now supplemented with a rate $4h^2\tau_0$. Similarly, all the gaps related to the Cooperons coupled by the valley-dependent SOC, $\Gamma_i^j$ $(i\in \{x,y\}$, $j \in\{0,z\}$, or  $i\in\{0,z\}$, $j \in\{x,y\})$, are supplemented with the rate $4\Delta_{so}^2\tau_0$. Furthermore, mixed terms of the form $\pm 4ih\Delta_{so}\tau_0$ introduce coupling of the Cooperons $C_z^{y(x)}$ with $C_x^{x(y)}$, and $C_y^{z(0)}$ with $C_0^{0(z)}$. }
 
\textcolor{black}{Next, let us consider the combinations of coupled Cooperons that enter the interference correction: $\mathcal{A}(_{s_1}^{l_1},_{s_2}^{l_2},_{s_3}^{l_3})=
	2\pi\nu\tau_0^2(C_{s_1}^{l_1}+C_{s_2}^{l_2}-C_{s_3}^{l_3})$, for $(_{s_1}^{l_1},_{s_2}^{l_2},_{s_3}^{l_3})=(_{z}^{y},_{x}^{x},_{0}^{x}),(_{z}^{x},_{x}^{y},_{0}^{y}),(_{y}^{z},_{0}^{0},_{x}^{0}),(_{y}^{0},_{0}^{z},_{x}^{z})$, as introduced in Eq.~\eqref{eqn25}. After inverting Eq.~\eqref{eqna1} and simplifying, we obtain
\begin{widetext}
\begin{equation}
\mathcal{A}(_{s_1}^{l_1}, _{s_2}^{l_2}, _{s_3}^{l_3})=\frac{
-\mathcal{P}_{s_1}^{l_1}\mathcal{P}_{s_2}^{l_2}+\mathcal{P}_{s_2}^{l_2}\mathcal{P}_{s_3}^{l_3}+\mathcal{P}_{s_3}^{l_3}\mathcal{P}_{s_1}^{l_1}+4\Delta_{so}^2(1+\mathcal{P}_{s_1}^{l_1}\tau_0+\mathcal{P}_{s_3}^{l_3}\tau_0+4\rho^2\tau_0^2)+4h^2(1+\mathcal{P}_{s_2}^{l_2}\tau_0+\mathcal{P}_{s_3}^{l_3}\tau_0+4\rho^2\tau_0^2)
	}{\mathcal{P}_{s_1}^{l_1}\mathcal{P}_{s_2}^{l_2}\mathcal{P}_{s_3}^{l_3}+4h^2 \mathcal{P}_{s_1}^{l_1}(1+\mathcal{P}_{s_2}^{l_2}\tau_0+\mathcal{P}_{s_3}^{l_3}\tau_0+4\rho^2\tau_0^2)+4\Delta_{so}^2\mathcal{P}_{s_2}^{l_2}(1+\mathcal{P}_{s_1}^{l_1}\tau_0+\mathcal{P}_{s_3}^{l_3}\tau_0+4\rho^2\tau_0^2)}.
\end{equation}
Finally, after neglecting terms which are small in the diffusive limit ($1\gg \mathcal{P}\tau_0, \rho^2\tau_0^2$), we find
\begin{equation}
\mathcal{A}(_{s_1}^{l_1},_{s_2}^{l_2},_{s_3}^{l_3})
=\frac{-\mathcal{P}_{s_1}^{l_1}\mathcal{P}_{s_2}^{l_2}+\mathcal{P}_{s_3}^{l_3}\mathcal{P}_{s_1}^{l_1}+4h^2+\mathcal{P}_{s_2}^{l_2}\mathcal{P}_{s_3}^{l_3}+4\Delta_{so}^2}
{\mathcal{P}_{s_1}^{l_1}\mathcal{P}_{s_2}^{l_2}\mathcal{P}_{s_3}^{l_3}+4h^2\mathcal{P}_{s_1}^{l_1}+4\Delta_{so}^2\mathcal{P}_{s_2}^{l_2}},
\end{equation}
\end{widetext}
which is exactly what enters Eq.~\eqref{eqn25}, where higher-order corrections were not included.}

\textcolor{black}{Thus, we have shown that the higher order corrections due to $\Delta_{so}$ and $h$ can be neglected. This result is not surprising, but it becomes apparent only at a late stage of the calculation, as it is contingent upon exact cancelation of several terms coming from two different sources: corrections to the Cooperon gaps of the form $4\Delta_{so}^2\tau_0$ and $4h^2\tau_0$, and the coupling of Cooperons by the terms of the form $\pm 4ih\Delta_{so}\tau_0$. This is a consequence of the basis chosen for our calculation.}

\textcolor{black}{\section{Diffusion constant in the regime $\tau_{iv}^{-1}\sim \tau_0^{-1}$} \label{ap2}}
 \textcolor{black}{We generalize the calculation of the transport time and the diffusion constant presented in Eq.~\eqref{eqn8}, to account for intra- and intervalley terms of the potential disorder $H_{\mathbf{qq'}}^{D0}$ on an equal footing. This yields
 \begin{equation} \tau_{tr}^{-1}=\tau_0^{-1}\frac{\mu^2+3E_g^2}{2(\mu^2+E_g^2)}+\tau_{iv,+}^{-1}+\tau_{iv,-}^{-1}+\frac{3}{2}\tau_{iv,x}^{-1}.
 \end{equation}	
Here, 
\begin{equation}
\tau_{iv,\pm}^{-1}=2\pi\nu\sum_{i=x,y}V_{\pm i}^2\bigg(1\pm \frac{E_g}{\mu}\bigg)^2
\end{equation}
describes on-site intervalley disorder, while 
\begin{equation}
 \tau_{iv,x}^{-1}=\pi\nu\sum_{i=x,y}V_{xi}^2\frac{v^2q_F^2}{\mu^2}
 \end{equation}
describes hopping intervalley disorder.
}

\textcolor{black}{
At $\mu\approx E_g$, the intervalley contribution to the transport time comes predominantly from one ("+") site, and the diffusion constant is
$D=\frac{1}{2}v_F (\tau_0^{-1}+\tau_{iv,+}^{-1})^{-1}$.
At $\mu\gg E_g$, both sites contribute equally, together with hopping disorder, and the diffusion constant is $D=\frac{1}{2}v_F (\frac{\tau_0^{-1}}{2}+\tau_{iv,+}^{-1}+\tau_{iv,-}^{-1}+\frac{3}{2}\tau_{iv,x}^{-1})^{-1}.$}


\begin{thebibliography}{40}
\bibitem [{foo()}]{foot1}%
\BibitemOpen
\href@noop {} {}\bibinfo {note} {The fact that there are 11 distinct
	scattering rates, excluding the dominant $\tau_0^{-1}$, can be understood as
	follows. As shown in Table \ref{table2} in Sec.~\ref{sec3}, there are 5
	distinct intervalley and 6 distinct intravalley gaps. Therefore, there should
	be 11 independent scattering rates to accommodate the same number of
	independent Cooperon channels.}\BibitemShut {Stop}%
\end{thebibliography}
\end{document}